\documentclass[preprint,article,accept,moreauthors,pdftex]{Definitions/mdpi} 

\firstpage{1} 
\pubvolume{xx}
\issuenum{1}
\articlenumber{5}
\pubyear{2020}
\copyrightyear{2020}
\history{Received: date; Accepted: date; Published: date}

\usepackage{tikz}
\usepackage{pgfplots}

\usetikzlibrary{arrows}

\usetikzlibrary{positioning}
\usetikzlibrary{arrows,fit}
\usetikzlibrary{shapes.geometric}
\usetikzlibrary{positioning,calc}
\usetikzlibrary{decorations.markings,shapes,arrows}

\tikzstyle{int}=[draw, fill=blue!10, minimum height = 1cm, minimum width=1.5cm,thick ]
\tikzstyle{sint}=[draw, fill=blue!10, minimum height = 0.5cm, minimum width=0.8cm,thick ]
\tikzstyle{sum}=[circle, fill=blue!10, draw=black,line width=1pt,minimum size = 0.5cm, thick ]
\tikzstyle{ssum}=[circle, fill=blue!10,draw=black,line width=1pt,minimum size = 0.1cm,inner sep=0pt]
\tikzstyle{int1}=[draw, fill=blue!10, minimum height = 0.5cm, minimum width=1cm,thick ]
\tikzstyle{enc}=[draw, fill=blue!10, minimum height = 2.7cm, minimum width=1cm,thick ]
\tikzstyle{int}=[draw, fill=blue!10, minimum height = 1cm, minimum width=1.5cm,thick ]

\usepackage{algorithm}
\usepackage{algorithmic}

\tikzstyle{int}=[draw, fill=blue!10, minimum height = .5 cm, minimum width=1 cm,thick ]
\tikzstyle{int1}=[draw,  minimum height = .15 cm, minimum width=1 cm,thick ]
\tikzstyle{tri}=[isosceles triangle, draw=black, fill=blue!10, minimum height = .1 cm, thin ]
\tikzstyle{sum}=[circle, fill=blue!10, draw=black ]	

\usepackage{amsmath, amssymb, bbm}
\usepackage{steinmetz,mathcomSTEv4}
\usepackage{xcolor}
\usepackage{enumitem}
\usepackage{subcaption}

\newcommand{\rr}{\mathrm {r}}

\newcommand{\ror}{\mathrm {r_{\zerov}}}
\newcommand{\rpar}{\mathrm {r_{p}}}

\newcommand{\ntq}{k}
\newcommand{\bps}{\rm bps}

\newcommand{\Acalo}{\overline{\Acal}}

\Title{
Comparison-limited Vector Quantization
}

\Author{Joseph Chataignon $^{1}$, Stefano Rini $^{2}$}

\AuthorNames{Joseph Chataignon, Stefano Rini}

\address{%
$^{1}$ \quad T\'el\'ecom Saint-\'Etienne, Universit\'e Jean Monnet,  France; joseph.chataignon@telecom-st-etienne.fr\\
$^{2}$ \quad Department of Electrical and Computer Engineering, National Chiao Tung University, Taiwan; stefano@nctu.edu.tw
}

\abstract{%
In this paper a variation of the classic vector quantization problem is considered.
{\color{black}
In the standard formulation, a quantizer is designed  to minimize the distortion between input and output when the number of reconstruction points is fixed.
We consider, instead, the scenario in which the number of comparators used in quantization is fixed. 
More precisely, we study the case in which a vector quantizer of dimension $d$ is comprised of $\ntq$ comparators, each receiving a linear combination of the inputs and producing the output value one/zero if this linear combination is above/below a certain threshold.
In reconstruction, the comparators' output is mapped to a reconstruction point, chosen so as to minimize a chosen distortion measure between the quantizer input and its reconstruction.
The Comparison-Limited Vector Quantization (CLVQ) problem is then defined as the problem of optimally designing the configuration of the compactors and the choice of  reconstruction points so as to minimize the given distortion.
In this paper, we  design a numerical optimization algorithm for the CLVQ problem.
This algorithm leverages combinatorial geometrical notions to describe the hyperplane arrangement induced by the configuration of the comparators. 
It also relies on a  genetic genetic meta heuristic to improve the selection of the quantizer initialization and avoid local minima encountered during optimization. 
We numerically evaluate  the performance of our algorithm in the case of input distributions following uniform and Gaussian i.i.d. sources to be compressed under quadratic distortion and  compare it to the classic Linde–Buzo–Gray (LBG) algorithm.
}
}

\keyword{
Constrained quantization; Source compression;  Hyperplane arrangement;  LBG algorithm; Gaussian quantizer; Uniform quantizer.
}

\begin{document}

\section{Introduction}
%

%
%
{\color{black}
Quantization represents one of the crucial operations in signal processing as it allows for the mapping of analog signals into discrete values, i.e. A2D conversion. 
Being such a fundamental processing operation, 
quantization has been an important subject of research and numerous results have been derived for problem.
In this work we consider a variation of the classic quantization setting  which appears to be so far neglected in the literature.
}
Vector quantizers are typically produced using operational amplifier (op-amp) comparators that take a signal and a bias as inputs and produce a zero/one voltage as output, depending on whether the signal is larger/smaller than the bias.
%
%
Generally speaking, op-amp comparators have high power consumption and manufacturing cost.
Additionally, generating a precise set of bias values  for each comparator gives rise to a rather complex circuit structure. 
{\color{black}
For this reason, in scenarios in which  complexity, fabrication cost, or  energy consumption  are severely restricted,   it is desirable to evaluate the  quantizer performance in terms of the number of comparators it requires to produce its output.}
This is in contrast with the classic approach, in which the quantizer performance is optimized under a constraint on the output cardinality, i.e. the number of bits used to represent the output in A2D conversion. 
%

\vspace{10cm}



\begin{figure}
	\centering
	\includegraphics[width=0.75\linewidth]{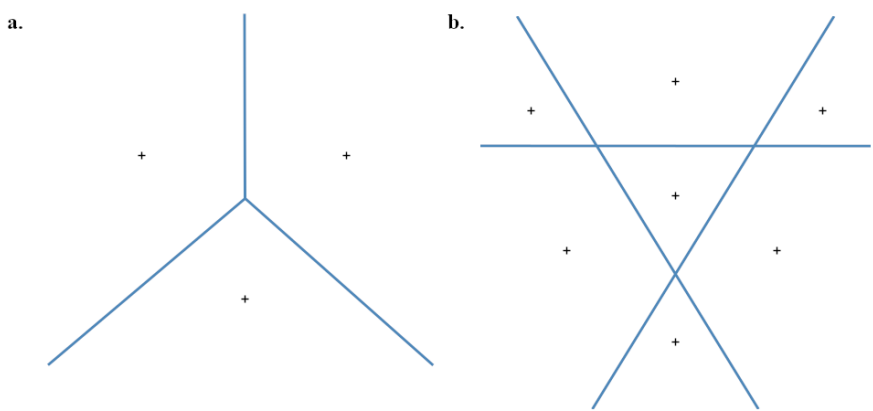}
\caption{Schematic representation of the vector quantization in which two input symbol are mapped into reconstruction points. 
In  (a), three reconstruction points are obtained using three comparators while, in (b) seven reconstruction points are obtained from three comparators
%
	    }
	\label{fig:quantization}
\end{figure}

{\color{black}
As a concrete example of the difference between the classic setting and the setting we propose, let us consider the case of a two dimensional quantizer ($d=2$) in which each dimension is quantized with a rate of $1.5$ bits-per-sample ($R=1.5 \ \bps$), so that 
In the classic setting, the quantizer is designed under the constraint that each pair of inputs is represented as one of three possible reconstruction points.
%
Under this constraint, a quantizer can be represented as in }
Fig. \ref{fig:quantization}: here each of the three reconstruction points is separated by the others through a edge in the Voronoi region, so that quantizer design requires three comparisons to distinguish the three reconstruction points. Accordingly, $\ntq=3$ comparisons are required to implement this quantization rule.
%
%
More generally, since every two reconstruction points are separated by an edge of the Voronoi region, a vector quantizer might require up to $2^R(2^R-1)/2$ comparators.
This scaling of the number of comparators with the output cardinality is generally valid only for low-rate regime, since  the number of neighbors of each reconstruction point is large.
Given that this quantization regime is naturally associated with low-cost and low-complexity devices, this cost scaling is particularly disadvantageous.
A question that naturally arises is whether a better scaling of the number of comparators with the output cardinality can be attained. 
To address this question, note  that number of comparators, $\ntq$, equals the total number of hyperplane segments in the Voronoi regions.
%
In this view, the best scaling is attained when the $\ntq$ hyperplanes induce the largest number of  partitions of the space of dimension $d$.
After some rather straightforward geometrical consideration, one realizes that for the case of $d=2$ and $R=1.5$, the largest number of partition is $7$.
This corresponds to the quantizer in Fig. \ref{fig:quantization}.
It is now apparent that there exists a large gap in the optimal quantizer design whether one considers a constraint on the number of points used in reconstruction or on the number of comparators employed.
{\color{black}
It is indeed the aim of this paper to develop
a first understanding of this problem, which we term the ``Comparison-Limited Vector Quantization (CLVQ) problem''.
}

%

\subsection{Relevant Results}
\label{literature}
Vector Quantization (VQ) \cite{gray1984vector,gersho2012vector} is a topic of vast interest, so that we shall briefly review only the literature which is most closely related to the topic of this publication.
Suffices to say that VQ is widely adopted in  speech coding \cite{makhoul1985vector,paliwal1993efficient,yahampath2007multiple}, image coding \cite{gersho1982image,nasrabadi1988image,chiranjeevi2018image}, and  video coding \cite{lee1995motion,regunathan2014adaptive}.
VQ has been  successfully used for speaker identification \cite{soong1987report,hasan2004speaker}, digital watermarking \cite{wu2005novel,lu2005multipurpose}, and clustering \cite{equitz1989new,lughofer2008extensions}.

In the following, we follow the  information theoretical study of the quantizer performance as described by rate-distortion theory \cite{berger2003rate}, originally developed by Shannon in  his  seminal work \cite{shannon1948mathematical}.
As originally formulated, the  \emph{rate distortion function} is the function describing the minimum rate at which a source of information can be compressed so that the reconstruction is to within a prescribed distortion from the original value, in the limit of infinitely many source samples are compressed. 
{\color{black}
This approach is  concerned with the compression rate needed to attain a prescribed distortion level, that is the  number of bits required on average to represent a source sample after quantization.}
The idea of accounting for the number of comparators required for A2D conversion emerges from the literature of mmWave MIMO communication  \cite{mezghani2007ultra}, in which high-resolution quantization is no longer feasible, duo to the high energy consumption in high-speed A2D conversion \cite{sundstrom2008power}.
For this reason, authors have investigated the  transmission performance  achievable when quantization only preserves the sign of the channel output \cite{mezghani2007ultra,mo2014high}. 
Generalizing the one-bit quantization case, in \cite{rini2017general} the authors investigate the capacity of a MIMO channel with output quantization constraint, i.e. the MIMO channel in which the channel output is processed at the receiver using a finite number of one-bit threshold quantizers. 
The channel model of \cite{rini2017general} is relevant in mm-wave communications which allows for a large number of receiver antennas, while the number of A2D conversion modules remains small due to limitation in the energy and costs of radio-frequency modules.
Building on an idea of \cite{mo2014high}, a connection between combinatorial geometry and the MIMO channel with output quantization constraints of \cite{rini2017general} is drawn in \cite{khalili2018mimo}.
In particular, in \cite{khalili2018mimo} it is shown that each quantizer can be interpreted as a hyperplane bisecting the transmitter signal space; for this reason  the largest rates are attained by the configuration allowing for the largest number of partitions induced by the set of hyperplanes.
To the best of our knowledge, the problem of hardware limited quantization has so far only been considered in \cite{shlezinger2018hardware}. 
The difference between our approach and that of  \cite{shlezinger2018hardware} is our focus on vector quantization, and our consideration of comparator limitations on the vector, rather than scalar, quantizer input.

Deep learning techniques have been used successfully in the domain of quantization, for task-based quantization problems \cite{shlezinger2019deeptask} as well as particular use cases (MIMO channel estimation in \cite{eldar2019deepquantization} or high-dimensional signal recovery from one-bit quantization in \cite{eldar2019deeprecovery}). The idea of adapting the quantizer design to the subsequent task, or task-based quantization, is explored in \cite{shlezinger2019quadratic} and studied in the context of hardware limitations in \cite{shlezinger2018hardware} \cite{shlezinger2019asymptotic}.
{\color{black}Tree-Structured Vector Quantization uses a set of scalar quantizers to gain speed, but comes at a disadvantage on the shape of quantization regions.}

Signal recovery from one-bit quantized signals is a topic previously investigated in the literature, such as in \cite{kamilov2012one} and \cite{plan2013one}.
An approach to one-bit signal recovery using deep network is considered in \cite{khobahi2019deep}.
These approaches can be interpreted as one-dimensional formulations of the CLVQ quantization problem we study here.
Algorithm for the design of quantizers under quadratic distortion is studied by  Max  and Lloyd for scalar quantization \cite{max1960quantizing,lloyd1982least}, and extended to vector quantization by Linde,Buzo, and Gray \cite{lindebuzogray1980vector}.

\subsection{Contribution}

{\color{black}
Our main contributions, in the following paper, consist in the definition of the Comparison-Limited Vector Quantization (CLVQ) problem in its full generality and identify the connections between this problem and the classic quantization  problem.
In particular, we highlight a  connection between the CLVQ problem and   combinatorial geometry  and provide novel insights on the ultimate performance of low-resolution quantization. 

We also provide a first class of algorithms for the CLVQ design: although not optimal, this first approach investigates the combinatorial geometric aspects of the CLVQ design. 
}
Numerical evaluations are provided for the case of uniform and Gaussian i.i.d. sources.  
The performance of the proposed quantizer is compared with the classic optimal vector quantizer design obtained by Linde-Buzo-Gray (LBG) algorithm \cite{lindebuzogray1980vector}.




\subsection{Paper Organization}
{\color{black}
The paper is organized as follows: Sec. \ref{sec:Related Results} presents related results in the literature and introduces 
some useful combinatorial notions.  Sec. \ref{sec:Vector Quantizer Model} presents the CLVQ model and the performance evaluation for this model. 
Sec. \ref{sec:Design Algorithm} introduces a class of  algorithms for the CLVQ problem. 
Sec. \ref{sec:Simulation results} provides relevant numerical evaluations for the case of an iid Gaussian source  to be reconstructed  under quadratic distortion. }
Finally, Sec. \ref{sec:Conclusion} concludes the paper.

\subsection{Notation}
In the remainder of the paper, all logarithms are taken in base two.
With $\xv= [x_1, \ldots, x_N] \subseteq \Xcal^N$ we indicate a sequence of elements from $\Xcal$ with length $N$.
The notation $\xv_i^j$ indicates the substring $[x_i, \ldots, x_j]$ of $\xv$.
The function $\sign(\xv)$ returns a vector with values in $\{-1,+1\}$ which equals the sign of each entry of the vector $\xv$.
%
%
%
%
The set $\{1,\ldots, N \}$ is indicated as $[N]$.
%
%
{\color{black}
Finally, the 2-norm operator is denoted as $\|\cdot \|$, expectation  as $\Ebb [\cdot]$, and $\onev(\cdot)$ is used for the indicator function.}

\section{Related Results}
\label{sec:Related Results}
{\color{black}
In this section let us review three topics of relevance in the remainder of the paper: (i) rate-distortion theory, that is the information theoretical study of the quantization, (ii) quantizer design algorithms, such as the  LBG algorithms which we use as a point of comparison of our results, and (iii) some combinatorial geometrical notions that are useful in describing quantizer configuration in the CLVQ problem. }

\subsection{Rate-Distortion Theory}
\label{sec:Rate-Distortion Theory}
The theoretical framework we introduce for the study of the CLVQ problem--see  Sec. \ref{sec:Vector Quantizer Model}-- expands on the classic formulation of rate-distortion theory \cite[Ch. 10 ]{cover1999elements}. 
Let us briefly review some relevant results in this section.

{\color{black}
Broadly speaking, rate-distortion theory is the information theoretical study of A2D conversion and  addresses the  quantization performance in the limit of an infinite source sequence when the number of bit-per-sample is kept constant. }

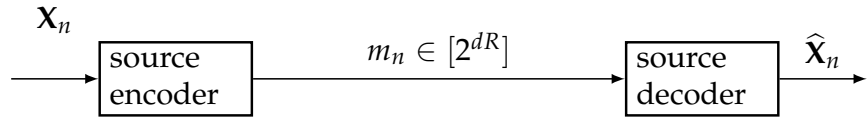
\begin{figure*}
	\centering
		\resizebox{0.75\textwidth}{!}{
			\begin{tikzpicture}[node distance=1cm,auto,>=latex]
				\tikzset{point/.style={coordinate},
					block/.style ={draw, thick, rectangle, minimum height=4em, minimum width=6em},
					line/.style ={draw, very thick,-},
				}
				\node  (d5) at (0,0) {};
				
				\node (d6) at (2,0)  [block,minimum height = 0.5 cm,minimum width=.5cm, text width=1.5cm] {source encoder};
				
				\draw (d5) [->,line width=0.5 pt]-- node[above, yshift=4mm]{ $\Xv_n$} (d6);
				
				\node (d7) at (8,0)  [block,minimum height = 0.5 cm,minimum width=.5cm, text width=1.5cm] {source decoder};
				
				\draw (d6) [->,line width=0.5 pt]-- node[above]{ $m_n \in [2^{dR}]$} (d7);
				
				\node (d8) at (10,0) {};
				
				\draw (d7) [->,line width=0.5 pt]-- node[above]{ $\Xhv_n$} (d8);

			\end{tikzpicture}
		}
	\caption{The rate-distortion problem in Sec. \ref{sec:Rate-Distortion Theory}.}
	\label{fig:system model RD}
\end{figure*}

Consider the system model in Fig. \ref{fig:system model RD}: 
%
here the  source sample $X^n=[X_{1},\ldots, X_{n}]$ with support $\Xcal^n$ is  represented through the index $m_n$ using the source encoder mapping 
\ea{
f_{\rm enc}: \Xcal^n \goes [2^{\lfloor n R \rfloor}],
\label{eq:enc}
}
so that the cardinality of $m_n$ is $2^{nR}$ when the source sequence has length $n$. 
The index $m_n$ is communicated to a source decoder through a noiseless channel. 
{\color{black}
In turn, the source encoder produces the reconstruction $\Xhv_n$  through the mapping
\ea{
f_{\rm dec}: \ [2^{\lfloor nR \rfloor}] \goes \widehat{\Xcal}^n,
\label{eq:dec}
}
so as to minimize the distortion measure $\rho_n(X^n,\Xh^n)$ between the original source sequence $X^n$ and its reconstruction $\Xh^n$.  
In \eqref{eq:dec},  $\rho_n$ for $n \in \Nbb$ is a positive and bounded function which captures the distortion, or degradation, between $X^n$ and $\Xh^n$.
The best attainable performance  when reconstructing the original symbol for a given value of $R$ is evaluated through the \emph{rate-distortion function} at rate $R$ and block-length $n$, defined as}
\ea{
D_n(R)=\inf  \rho_n(X^n,\Xh^n),
\label{eq:opt RD}
}
where the infimization is over all source encoder/decoder mappings as in \eqref{eq:enc}/\eqref{eq:dec} .
%
%
One is generally interested in the performance when $n$ goes to infinity, since it provides an upper bound to the ultimate compression performance. 
We refer to this limit as the \emph{rate distortion function} at rate $R$:
\ea{
D(R)= \liminf_{n \goes \infty} \ \  D_n(R).
\label{eq:distortion rate function}
}
It is well known that,  when (i) the source sample is i.i.d. from the distribution $P_X$ and (ii) the distortion function is an additive distortion function, i.e.
\ea{
\rho_n(X^n;\Xh^n)= \sum_{i \in [n]} \rho(X_i,\Xh_i),
\label{eq:n distortion}
}
{\color{black}
for some positive and bounded $\rho(x_i,\xh_i)$, we have }
\ea{
D(R) = \inf_{P_{\Xh|X}, \ R {\geq} I(X,\Xh)} \  \Ebb \lsb \rho(X,\Xh) \rsb,
\label{eq: DR}
}
where $I(X,\Xh)$ indicates the \emph{mutual information} between the source $X$ and the reconstruction $\Xh$.
As an example of the result in \eqref{eq: DR} consider the case in which $X$ has a standard normal distribution to be reconstructed under quadratic distortion,  then
\ea{
D(R)= 2^{-2R},
}
so that  $D \in [0,1]$.

\subsection{Quantization Design Algorithms}
\label{sec:Quantization Algorithm}

{\color{black}
The approach in Sec. \ref{sec:Rate-Distortion Theory} 
%
provides the theoretical limit to the quantization performance.
%
In order to approach such performance, various algorithms have been proposed in the literature for the design of high-dimensional quantizers. 

In the following, we shall compare the performance of the CLVQ design with the standard LBG algorithm which we shall describe next.
The LBG 
algorithm, also known as the  Max-Lloyd algorithm or $k$-means clustering algorithm, is an iterative algorithm for the  design of quantizers for the case of quadratic distortion \cite{max1960quantizing,lloyd1982least,lindebuzogray1980vector}. 
It is known to approach the optimal solution although this problem is NP-hard in the Euclidean space. 
The  pseudo-code for the algorithm is provided in Algo. \ref{al:quantization}: the algorithm performs an alternate optimization between the reconstruction regions and the position of the reconstruction points. 
In the first step, the sample space $\Xcal$ is partitioned by a Voronoi tessellation with the reconstruction points as generators. This means that each point in the sample space is assigned to the closest reconstruction points.
reconstruction point has a region, and every point in space belongs to the region of the nearest reconstruction point.
This procedure results in a set of regions $\{\Vcal_m\}_{m \in [M]}$  so that each point in the region $\Vcal_m$ contains the points that have the minimum distortion when reconstructed as the point $\ch_m$, instead of any other reconstruction point. 
%
In the second step,  the position of the reconstruction points $c_m$ is optimized: each reconstruction point $c_m$ is chosen as the mean of the sample points in the region $\Vcal_m$.
%
%
The two steps are repeated in succession until convergence is achieved. 
Various procedures can be designed to initialize the initial positions of the set of reconstruction points $\{\ch_m\}_{m \in [M]}$
as it is known that the choice of initialization set its crucial to attain fast convergence. 

In the literature, many variations of the  LBG algorithm as been studied with applications ranging from unsupervised learning to low-ranking approximations.
Let us mention two variations of interest as a generalization of the settings in this paper.

Entropy-constrained vector quantizers (ECVQs) are quantizers in which the quantization objective is the distortion minimization subject an additional a bound on the entropy of the quantizer output. 
ECVQ are useful when they are used in tandem with variable-rate noiseless coding systems to provide locally optimal variable-rate block source coding.
Another class of quantizers of interest are lattice quantizers. In lattice quantizers are quantizers in which the set of reconstructions points form a lattice. 
Note that the problem of determining the closest reconstruction point to a given quantizer input is, in general, a nearest neighbor problem. 
This class of problems has a complexity linear in the number of neighbors: when a quantizer has high dimension, this complexity is not amenable to real-time applications. 
For this reason, one is often interested in introducing further structure in the set $\{\ch_m\}_{m \in [M]}$ with the aim of speeding up the quantization operation. 
One class of such quantizers are lattice vector quantizers, that is, quantizers in which the set of reconstruction points form a lattice.
It has been shown that lattice quantizers have good performance while also allowing for a fast quantization \cite{conway1982fast,zamir1992universal}.
 }
 




\begin{algorithm}
\caption{Max-Lloyd/$k$-means clustering/Linde-Buzo-Gray (LBG) Algorithm}
\begin{spacing}{1.5} 
\begin{algorithmic}[1]
\STATE {\bf Input:} source distribution $P_{X^n}$, number of reconstruction points $M=2^{\lfloor n R \rfloor}$, maximum number of iterations $T_{\max}$
\STATE {\bf Output:} position of the reconstruction points $\{\ch_m\}_{m \in [M]}$, Voronoi region of each reconstruction point $\{\Vcal_m\}_{m \in [M]}$
\STATE randomly pick the initial reconstruction points  $\{\ch_m(0)\}_{m \in [M]}$, i.e.
\ean{
\ch_m(0) \sim P_{X^n}
}
\FOR{$t$ in $[T_{\max}]$}
\STATE  update the Voronoi regions $\{\Vcal_m(t)\}_{m \in [M]}$ as
\ean{
\Vcal_m(t)= \lcb x \in \Xcal \ST  (c_m(t-1)-x) ^2 \leq (c_j(t-1)-x)^2,  \ \forall j \in [M]\setminus m \rcb 
}
for all {$m \in [M]$}
\STATE update the position of the reconstruction points  $\{c_m(t)\}_{m \in [M]}$ as
\ean{
c_m(t)= \f {\Ebb [X | X \in \Vcal_m(t)]}{ \Ebb [1_{ \{X \in \Vcal_m(t)} \}]}
}
\ENDFOR
\end{algorithmic}
\end{spacing}
\label{al:quantization}
\end{algorithm}

{
\subsection{Some combinatorial geometric notions}
\label{sec:Combinatorial Interlude}
}
%
%
This section briefly introduces a few combinatorial concepts useful in the remainder of the paper to describing the geometrical properties of the proposed quantization scenario.
{\color{black}
In this section and through the paper we mostly use the nomenclature and notation of \cite{stanley2006}
}
A  \emph{hyperplane arrangement}  $\Acal$ is a finite set of $n$ affine hyperplanes in $\Rbb^m$ for some $n,m \in \Nbb$.
%
where $\Av$ is obtained by letting each row $i$ equals $\av_i^T$ and letting $\bv=[b_1 \ldots b_n]^T$.
A \emph{region}, $\Rcal_i$ of an arrangement $\Acal$ is a connected component of the complement $\Acalo$ of the hyperplanes, defined as 
\ea{
\Acalo= \Rbb^m - \Acal.
}
Let $\rv(\Acal)$ be the number of regions in which the hyperplane arrangement $\Acal$ divides the space $\Rbb^m$, so that
\ea{
\Acalo = \bigcup_{i=1}^{\rv(\Acal)} \Rcal_i.
}
%
A plane arrangement $\Acal$ is said to be in General Position (GP) if and only if:
$$\{ H_1, ... ,H_p \} \subseteq \Acal , p \leq n \Rightarrow dim(H_1 \cap ... \cap H_n) = n-p $$
$$\{ H_1, ... ,H_p \} \subseteq \Acal , p > n \Rightarrow H_1 \cap ... \cap H_n = \emptyset $$

%
%
%
\begin{Lemma}{\bf \cite[Th. 1.2]{dimca2017hyperplane}.}
	\label{lem:hyperplane}
	A hyperplane arrangement of size $n$  in  $\Rbb^m$ divides $\Rbb^m$ into at most $\rr(m,n)$ region for
	\begin{equation}
	\rr(m,n)=
	\sum_{i=0}^{m} {n \choose i}\leq 2^n.
	\end{equation}
	%
\end{Lemma}
\begin{Lemma}{\bf \color{black} \cite[Prop. 2.4]{stanley2006}.}
	\label{lem:origin arrangement}
	A hyperplane arrangement of size $n$  in  $\Rbb^m$  where all the hyperplanes pass through the origin 
	divides $\Rbb^m$ into at most  $\ror$ regions for 
	\ea{
		\ror(n,m)= 2\sum_{i=0}^{m-1} {{n-1}\choose{i}}.
	}
\end{Lemma}

\begin{Lemma}{\bf \color{black} \cite[Prop. 2.4]{stanley2006}.}
	\label{lem:pcomp}
	Let $\Acal$ be a hyperplane arrangement of size  $l$ in $\Rbb^m$ and consider a hyperplane arrangement $\Bcal$ of size $dl$ with $d \in \Nbb$ hyperplanes parallel to each of the hyperplanes in $\Acal$, then $\Bcal$ 	divides $\Rbb^m$ into at most $\rpar$ regions for 
	\begin{equation}
	\rpar(m,n,d)=
	\sum_{i=0}^{m} \binom{l}{i} d^i  \leq (1+d)^l.
	\label{eq:pcomp}
	\end{equation}
	%
\end{Lemma}
A necessary condition to attain the equality in Lem. \ref{lem:hyperplane}, \ref{lem:origin arrangement},  and Lem. \ref{lem:pcomp} 
is for the hyperplane arrangement $\Acal$ to be in GP.
As we shall see in the next section,

\section{Comparison-Limited Vector Quantization (CLVQ) Problem}
\label{sec:Vector Quantizer Model}

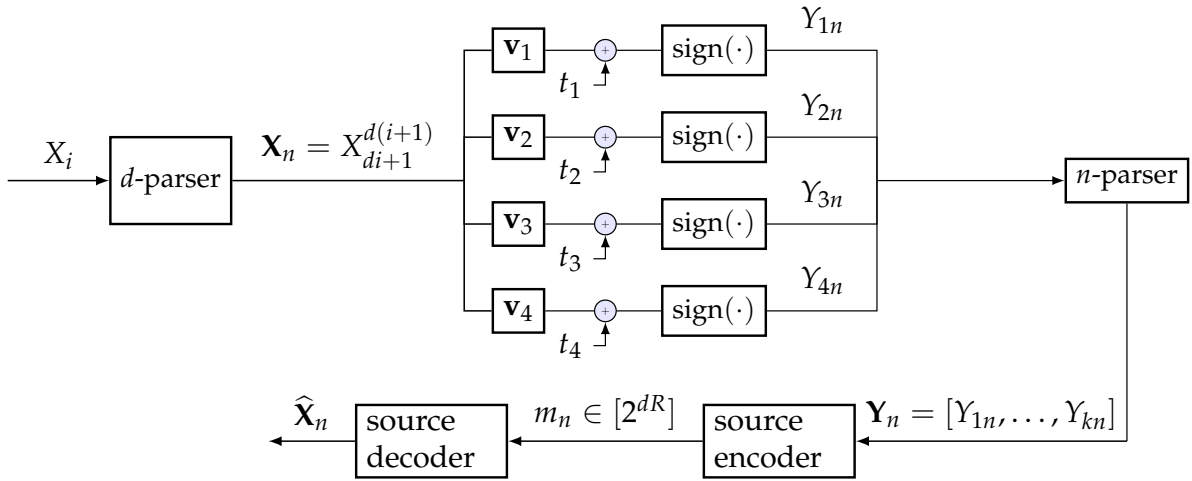
\begin{figure*}
	\centering
	\resizebox{1.0\textwidth}{!}{
		\begin{tikzpicture}[node distance=2.5cm,auto,>=latex]
		\tikzset{point/.style={coordinate},
			block/.style ={draw, thick, rectangle, minimum height=4em, minimum width=6em},
			line/.style ={draw, very thick,-},
		}
		\node (b1) at (0,0){};
		\node (b2) [block, minimum height = 1 cm, minimum width=1cm,thick =1cm] at (2,0) {\small $d$-parser};
		%
		\draw (b1) [->,line width=0.5 pt]-- node[above]{ $X_i$} (b2);

		\node (d1) at (6,1.5)  [block,minimum height = 0.5 cm,minimum width=.5cm] {  $\vv_1$};
		\node (d2) at (6,0.5)  [block,minimum height = 0.5 cm,minimum width=.5cm] {  $\vv_{2}$};
		\node (d3) at (6,-0.5)  [block,minimum height = 0.5 cm,minimum width=.5cm] {  $\vv_{3}$};
		\node (d4) at (6,-1.5)  [block,minimum height = 0.5 cm,minimum width=.5cm] {  $\vv_{4}$};

		\node (b3) at (5.5,0){};
		
		\draw (b2) [line width=0.5 pt]-- node[above]{ $\Xv_n=X_{di+1}^{d(i+1)}$} (b3);
		
		\draw (b3.west) [line width=0.5 pt]|- (d1);
		\draw (b3.west) [line width=0.5 pt]|- (d2);
		\draw (b3.west) [line width=0.5 pt]|- (d3);
		\draw (b3.west) [line width=0.5 pt]|- (d4);
		
		\node  at (7,-1.5) (e11) [sum,scale=0.35]{$+$};
		\node  at (7,-.5) (e21) [sum,scale=0.35]{$+$};
		\node  at (7,+.5) (e31) [sum,scale=0.35]{$+$};
		\node  at (7,+1.5) (e41) [sum,scale=0.35]{$+$};

		\node  at (7-0.4,-1.5-0.4) (th1) {   $t_4$};
		\node  at (7-0.4,-.5-0.4) (th2) {$t_3$};
		\node  at (7-0.4,+.5-0.4) (th3) {   $t_2$};
		\node  at (7-0.4,+1.5-0.4) (th4) {   $t_1$};
		
		\node  at (8.25,-1.5) (quant1) [int1] {\small  $\sign(\cdot)$};
		\node  at (8.25,-.5) (quant2)  [int1] {\small  $\sign(\cdot)$};
		\node  at (8.25,+.5) (quant3)  [int1] {\small  $\sign(\cdot)$};
		\node  at (8.25,+1.5) (quant4) [int1] {\small  $\sign(\cdot)$};
		
		\draw (th1) [->]-| (e11);
		\draw (th2) [->]-| (e21);		
		\draw (th3) [->]-| (e31);
		\draw (th4) [->]-| (e41);
		
		\draw (d1) -- (e41);
		\draw (d2) -- (e31);		
		\draw (d3) -- (e21);
		\draw (d4) -- (e11);
		
		\draw (quant4) -- (e41);
		\draw (quant3) -- (e31);		
		\draw (quant2) -- (e21);
		\draw (quant1) -- (e11);
		
		\node (d5) at (13,0)  [block,minimum height = 0.5 cm,minimum width=.5cm] {\small $n$-parser};
		
		\node (d6) at (10,0){};
		
		\draw (quant4)-| (d6.east);
		\draw (quant3)-| (d6.east);
		\draw (quant2)-| (d6.east);		
		\draw (quant1)-| (d6.east);	
		\draw (d6)[->]-- (d5.west);								
		
		\node  at (9.5,1.5+0.35)  {  $Y_{1n}$};
		\node  at (9.5,.5+0.35)  {  $Y_{2n}$};
		\node  at (9.5,-.5+0.35)  {  $Y_{3n}$};
		\node  at (9.5,-1.5+0.35)  {  $Y_{4n}$};
		
        \node (d7) at (9,-3)  [block,minimum height = 0.5 cm,minimum width=.5cm, text width=1.5cm] {source encoder};
		\node (d8) at (13,-3) {};
		\draw (d8.center) [->,line width=0.5 pt]-- node[above]{  $\Yv_n=[Y_{1n},\ldots,Y_{\ntq n}]$} (d7);
		
		\draw (d5) -- (d8.center);
		
 		\node (d9) at (5,-3)  [block,minimum height = 0.5 cm,minimum width=.5cm, text width=1.5cm] {source decoder};
		
 		\node (d10) at (3,-3) {};
		
 		\draw (d7) [->,line width=0.5 pt]-- node[above]{ $m_n \in [2^{dR}]$} (d9);
		
 		\draw (d9) [->,line width=0.5 pt]-- node[above]{ $\Xhv_n$} (d10);

		\end{tikzpicture}
	}
	\vspace{0.25 cm}
	\caption{The Comparison-Limited Vector Quantization (CLVQ) problem with $\ntq=4$ quantizers.}
	\label{fig:system model}
\end{figure*}

{\color{black}
After the introductory notion in Sec. \ref{sec:Related Results}
let us introduce the Comparison-Limited Vector Quantization (CLVQ) problem, also conceptually depicted in Fig. \ref{fig:system model}.
}

%

A source sequence $\{X_i\}_{i\in \Nbb}$, where each $X_i$ has support $\Xcal$, 
%
is parsed in super-symbols $\{\Xv_n\}_{n\in \Nbb}$ of dimension $d$ with $\Xv_n=[X_{d(n-1)+1}, \ldots, X_{dn}]$ for $n \in \Nbb$ where $d$ is referred to as the \emph{dimension} of the vector quantizer.
The $j^{\rm th}$ comparator computes a linear combination of each super-symbol $\Xv_n$ and outputs a signal $Y_{jn}$  as
\ea{
Y_{jn} = \sign\lb \vv_j \Xv_n+t_j \rb,
\label{eq:Yjn}
}
for $j \in [\ntq]$. 
The value $\ntq$ is called \emph{resolution} of the vector quantizer;
$\vv_j \in \Rbb^{d}$, $t_j \in\Rbb$ are fixed and known. 
For more convenience, we express the $\ntq$ outputs of the quantizers of \eqref{eq:Yjn} as
\ea{
\Yv_n=\sign \lb \Vv \Xv_n+\tv \rb,
\label{eq:Yv n}
}
where $\Vv= \in \Rbb^{\ntq \times d}$ is such that its $i^{\rm th}$ row is equal to the vector $\vv_i$ from \eqref{eq:Yjn}.  
Similarly, $\tv \in \Rbb^{\ntq}$ has the  $i^{\rm th}$  entry equal to $t_i$ and  $\Yv_n=[Y_{1n},\ldots,Y_{\ntq n}]$.
The set $[\Vv,\tv]$ is referred to as the configuration of the quantizer.
%
%
The super-symbol $\Yv_n$ is provided to a source encoder 
%
%
that produces a bit-restricted representation of the quantizers' output as $m_n \in [2^{\lfloor dR \rfloor}]$ where $R$ is the name given to the \emph{rate} of the quantizer through the \emph{source encoding mapping}
\ea{
f_{\rm enc}: \ \{-1,+1\}^{\ntq} \goes [2^{\lfloor dR \rfloor}].
\label{eq:source encoder}
}
The message $m_n$ is sent into the source decoder which outputs a reconstruction of the source super-symbol $\Xv_n$ as $\Xhv_n= [\Xh_{dn+1}, \ldots, \Xh_{d(n+1)}]$ 
with $ \Xh_i \in \widehat{\Xcal}$ ,  through the \emph{source decoding mapping}
\ea{
f_{\rm dec}: \ [2^{\lfloor dR \rfloor}] \goes \widehat{\Xcal}^d.	
\label{eq:source decoder}
}
We measure the effectiveness of a vector quantizer with a measure of distortion
\ea{
\rho_n(X^n;\Xh^n): \ \Xcal^n \times \widehat{\Xcal}^n \goes \Rbb^+,
} 
for  $n	\in \Nbb$ which is assumed positive and non-decreasing in $n$.
For a given configuration of the linear combiners $[\Vv,\tv]$, source encoder/decoder mappings $f_{\rm enc}$/$f_{\rm dec}$,  and 
given a distortion measure between input and reconstruction sequence $\rho_n(X^n;\Xh^n)$,  the performance of the quantizer is evaluated as
\ea{
\overline{\rho}= \limsup_{n \goes \infty} \  
\rho_n(X^n;\Xh^n).
\label{eq:distortion}
}
The optimal quantizer performance regarding the distortion $\overline{\rho}$, dimension $d$, resolution $\ntq$ and rate $R$ can be written as 
\ea{
D(d,\ntq,R)=\inf \ \ \overline{\rho},
\label{eq:opt}
}
where the infimization is over all  configuration of the linear combiners $\textit{\textbf{[V,t]}}$ and all the source encoder/decoder mappings.
%

%
{\color{black}
Similarly to \eqref{eq:distortion rate function}, one is often interested in determining the limiting performance as the dimension of the quantizer grows to infinity. 
In this case, it is natural to let the number of quantizer grow exponentially with the dimension, so that the number of  comparators per dimension is kept constant, with ration approximatively equal to $\al$.
}
Let $\ntq= \lceil \al d \rceil$,  then the comparison-limited  distortion-rate function is defined as 
\ea{
D_{\al}(R)=\lim_{d \goes \infty, \ \ntq= \lceil \al d \rceil }D(d,\ntq,R),
\label{eq: D R al}
}
for $D(d,\ntq,R)$ in \eqref{eq:opt}, that is $D_{\al}(R)$ is the minimum distortion attainable as the quantizer dimension grows to infinity while $k/d\approx \al$.
%
%

{\color{black}
\subsection{Underlying Assumptions}
The underlying assumptions in our problem formulation are as follows:
op-amp voltage comparators are employed in nearly all analog-to-digital converters to obtain multilevel
quantization.
A reconfigurable receiver front-end might be able to re-configure the comparators' inputs so as to perform more complex operations.
%
%
%
Given the receiver's ability to partially reconfigure its circuitry depending on the channel realization, we wish to determine which
configuration of the comparators  yields the largest capacity.
%
%
We restrict our results to the case of linear analog pre-coding, as the capacity of the model without output quantization constraints can be attained through linear pre-coding strategies.
The extension to non-linear processing is possible but not pursued here.
}

\subsection{Discussion}
\label{sec:Discussion}

Let us connect the CLVQ problem as defined above  with the results in Sec. \ref{sec:Related Results}. 
As it can be gathered by comparing Fig. \ref{fig:system model RD} versus Fig. \ref{fig:system model}, the CLVQ problem explicitly embeds a model of the hardware architecture performing A2D.
%
{\color{black}
In particular, in this formulation, we assume that quantization is performed through a bank of op-amp amplifier. Each op-amp amplifier receives a linear combination of the quantizer input and a bias.
}

\medskip

{\color{black}
It is immediate to connect the problem formulation in Fig. \ref{fig:system model}  with the setting in Sec. \ref{sec:Combinatorial Interlude}. 
The set of  the comparators output $\Yv_n$ in \eqref{eq:Yv n} represent the membership function of each quantizer input $\Xv_n$ with respect to each of the hyperplanes $[\vv_i, t_i]$. 
This output is then compressed as in the classic rate-distortion problem in Sec. \ref{sec:Rate-Distortion Theory}. 
When the number of sign quantizers is sufficiently large, a hyperplane can be used to separate any pair of reconstruction points, as in the Voronoi region of Algo. \ref{al:quantization}. 
In this regime, the rate in \eqref{eq: DR} is determined by the number of regions induced by the hyperplane arrangement. 
%
%
Given the considerations above, it is now useful to imagine the CLVQ problem in the context of Fig. \ref{fig:quantization}: each comparator induces a hyperplane partitioning the sample space. The set of comparators then can be imagined as a hyperplane arrangement partitioning the sample space. The set of comparator outputs is then compressed as in the classic rate-distortion problem, thus further reducing the number of bits required to represent $\Yv^n$ by exploiting its distribution.  

\medskip

Let us next provide some more specific remarks.

\begin{Remark}{\bf Extension to the vector version}
In the following we consider the case in which the source sequence $\{X_i\}_{i \in \Nbb}$ is scalar. The case of a vector input sequence is not considered here. 
For the vector case, the dependency among different dimensions can be addressed in different manner: two come readily to mind.
One approach would be to choose the linear combination as the whitening transformation that make the covariance of the input symbol  unitary.
Alternatively, the different dimension can be processed in parallel by the comparator network while the source encoder  is tasked with accounting for the correlation across dimensions.
\end{Remark}

\begin{Remark}{\bf Entropy coding:}
\label{rem:Entropy coding}
Note that the CLVQ problem formulation also captures the digital processing following  A2D conversion. 
After quantization, the quantized samples are further compressed using a digital source compressor, such as an entropy coder.  This is captured by the source encoder/decoder in \eqref{eq:source encoder}/\eqref{eq:source decoder} and as in Fig. \ref{fig:system model}. 
This is similar to the ECVQ discussed in Sec. \ref{sec:Quantization Algorithm}.
%
%
\end{Remark}

\begin{Remark}{\bf Limitations in the comparator configurations:}
In \eqref{eq:opt} we assume that the CLVQ performance can be optimized over all possible   configuration of the linear combiners $[\Vv,\tv]$.
As a configuration of a linear combiner corresponds to an analog circutry implementation, this assumption might be unfeasible in some scenarios, as generating precise linear combinations and voltage references might require a high circuit complexity.
To accommodate for such further limitations, on can further restrict the optimization in \eqref{eq:opt} over a set of configurations which can be efficiently implemented.
Such an approach is similar to that in \cite{rini2017general} and further discussion on this approach can be found there.
\end{Remark}

\begin{Remark}{\bf Lattice quantization for CLVQ:}
Note that one can extend the construction of lattice codes for quantization to the CLVQ problem by considering regular hyperplane arraignments. 
In this case, it is possible to similarly define computationally-efficient quantization algorithms as discussed in Sec. \ref{sec:Quantization Algorithm}.

\end{Remark}
}

\section{A CLVQ Quantizer  Design Algorithm}
\label{sec:Design Algorithm}
{\color{black}
In the following, we shall introduce an algorithm for the design of the  CLVQ linear  combiner configuration $[\Vv, \tv]$ for given value of $d$, $k$, and $R$ in \eqref{eq:opt}. 
In other words, the algorithm attempts to numerically determine the optimal solution to \eqref{eq:opt} by assuming that the optimal choice of source encoder/decoder function is performed in a successive optimization step. 
}

%
For the design of this numerical optimization algorithm, we take inspiration from the quantizer design algorithms discussed in Sec. \ref{sec:Quantization Algorithm}, although the CLVQ problem is more general than considered is Sec. \ref{sec:Quantization Algorithm}.
Indeed, we shall leverage the notion in Sec. \ref{sec:Combinatorial Interlude} to more efficiently represent and manipulate the linear combiner configuration during the optimization. 
The theoretical analysis of the asymptotically optimal solution as in \eqref{eq: D R al} is left for future research. 

\medskip
{\color{black}
For simplicity, in the design of our algorithm, we simplify the general setting of Sec. \ref{sec:Vector Quantizer Model} as follows:
}
\begin{enumerate}[label=(\alph*)]
    
\item 
{\bf i.i.d. sources --} with distribution unimodal distribution $P_X$, 

\item 
{\bf quadratic distortion --} also known as Mean Squared Error (MSE) distortion, i.e. 
\ea{
\rho_n(X^n;\Xh^n)=\f 1 n \sum_{i=1}^n \|X_i-\Xh_i\|^2.
\label{eq:mse}
} 
\item 
{\bf infinite compression rate --} that is  $R=\infty$ in  \eqref{eq:opt}, i.e.  $D(d,\ntq,\infty)$. 
\end{enumerate}

{\color{black}
%
Regarding assumption (b), this assumption is relaxed  in Sec. \ref{sec:entropy_maximization}, here we discuss the use of entropy as a measure of distortion measure in the numerical optimization.  
%
%
%
Assumption (c) is chosen so as to simply the design of the comparator configuration.
An effective, but sub-optimal, approach to address this constraint would be choose the encoding and decoding function which implement an efficient lossy-compression algorithm for discrete data -- see \cite{nelson1995data}. 

Note that a more efficient approach to address the case $R<\infty$  would be to consider an entropy constrain, as discussed in Rem. \ref{rem:Entropy coding}.
}
%
The design of such an algorithm would be similar to the problem of entropy-constrained  quantization as in \cite{chou1989entropy}. This further design step  is not considered at the present. 


%
%
%

\medskip

Given the three assumptions above, \eqref{eq:opt} simplifies to
\ea{
D(d,k,\infty)=
\sum_{i=1}^d \Ebb \lsb \|X_i -\Xh_i \|^2 \rsb,
\label{eq:simply to expect}
}
where $\Xv=[X_i,\ldots,X_d]$ and $\Xhv=[\Xh_1,\ldots,\Xh_d]$ for $X_i,\Xh_i $ in \eqref{eq:simply to expect} are i.i.d. distributed, so that the subscript $n$ can be dropped.\footnote{That is, every super-symbol $\Xv_i$ is quantized in the same manner, regardless of $n$.}
For the problem in \eqref{eq:simply to expect}, the linear combiner optimization problem is mapped to a combinatorial geometric optimization problem as in Sec. \ref{sec:Combinatorial Interlude} in which one is interested in choosing the hyperplane arrangement which  divides the space in $\{\Rcal_i\}$  regions for which MSE distortion
{\color{black}
is minimized as in Algo. \ref{al:quantization} for the case in which  each  source vector $\Xv_n$ is mapped to the closest centroid in the set $\{c_i\}_{i \in [\rv(\Acal)]}$.
}
The set of centroids is obtained from the hyperplane arrangement $\Acal$ as 
 \ea{
c_i=\f {\Ebb [X | X \in \Rcal_i]}{ \Ebb [1_{ \{X \in \Rcal_i} \}]},
}
similarly to the LBG algorithm. 

 
\medskip

Let us further detail the proposed approach: our  algorithm is divided in two steps:  (i) an initialization step --discussed in Sec. \ref{sec:Initialization methods}
-- and (ii) an optimization step -- detailed in Sec. \ref{sec:Optimization step}.

Generally speaking, the optimization step is computationally complex, so the role of the initialization step is to choose of a set of initial configurations of quantizer that can span the large set of choices for the combiner configuration. 
By considering multiple restarts of the optimization point with a different initialization, one hopes to converge to an optimal solution and avoid local minima.

Note that there is a large number of symmetries in the linear combiner configurations, since one can permute each quantizer configuration and input values and obtain an equivalent configuration. 
This induces a number of local minima which hinder the convergence of the optimization step to the globally optimal solution. 
For this reason, the choice of the set of combiner configuration used for initialization  is crucial to span the space of possible solutions.

\subsection{Initialization methods}
\label{sec:Initialization methods}

In the following, we explore two possibilities for initialization:

\begin{description}
    \item {\bf -- random initialization:} in which the set of initial linear combiner configurations selected  using random samples from the source distribution. 
    \item {\bf -- genetic initialization:} in which we utilize a genetic algorithm to generate the set of initial configurations.
\end{description}


\subsubsection{Random initialization}
\label{sec:Random initialization}
This first approach to the generation of initialization steps is the simplest, but yet provides good results.
To obtain random hyperplanes that cut through dense zones of the source distribution, it is sufficient to randomly generate $d$ points  following that same source distribution and the hyperplane passing through these points can be used as a linear combiner configuration. By repeating this operation $\ntq$ times, one obtains an initial configuration for every comparator.
This random initialization  has constant time complexity and allows us to quickly generate a large number of possible candidate arrangements.

\subsubsection{Genetic Initialization}
\label{sec:genetic initialization}

Given the poor performance of the random initialization in Sec.  \ref{sec:Random initialization}
and the computational complexity of the exhaustive listing of all possible configuration in App. \ref{sec:Progressive arrangement growth}, we next consider a  genetic algorithm that retains elements of the two previous approaches. 
%
%
A genetic algorithm is a meta-heuristic taking inspiration from genetic evolution to apply some processes such as selection, crossover and mutation to optimization problems.
It consists of a few steps that are repeated over several iterations.
An initial population of potential solutions has to be generated. At each iteration, a part of the current generation is selected to breed the next one, based on a fitness function measuring its efficiency regarding the problem.
Then, another generation is formed by performing crossover (which corresponds to mixing some elements between pairs of solutions) and mutation (small alterations of solution) on the current generation.
%

In our case, after generating a few starting configurations according to the random approach in Sec. \ref{sec:Random initialization}, we proceed to genetic selection in three distinct stages:
%
\begin{enumerate}[label= (\alph*)]
	\item select a set of configurations based on their fitness (the MSE in our case)
	\item perform crossover on these configurations to obtain the next generation with improved fitness
	\item apply random mutations to the resulting configurations
\end{enumerate}

\noindent
$\bullet \ $  In step (a),  the configurations are reordered by increasing distortion (we use the opposite of distortion as fitness function).The $k$ first configurations are kept for the next step, $k$ being a fixed parameter.
%
In our simulations, we set $k$ to $80\%$ of the number of configurations.

%

\noindent
$\bullet \ $ In step (b), we cross pairs of arrangements following one of two policies.
%
The first one consists in randomly selecting hyperplanes to form pairs.
With the second policy, we use a dissimilarity value to form the pairs of hyperplanes.
This dissimilarity value is 
given by $\theta \times d(p_1,p_2)$, where $\theta$ is the angle between the hyperplanes; $d$ is the euclidean distance; and $p_1,p_2$ are the nearest point of each hyperplane to the centroid of the source distribution.
Measures of the dissimilarity of each pair of hyperplanes are placed in a matrix, and hyperplanes are associated two by two in a way that produces the lowest average dissimilarity. 
%
Once pairs are formed, a new configuration is formed by selecting one hyperplane from each pair.

\noindent
$\bullet \ $ 
Step (c) consists in mutating the resulting configurations.
%
To apply mutation, we first form vectors of random numbers following a normal probability law with mean $\mu=1$ and with a standard deviation $\sigma = 0.2$ that was empirically chosen. Mutation is only applied to the half of configurations that have the lowest fitness score. The intuition for applying mutation this way is that solutions with a good fitness value might be made worse  by a mutation, while the ones which fitness is already not good have better chances of being improved by the mutation.

\medskip
These steps are repeated for a pre-determined number of iterations. Each step has a constant complexity, so the complexity of each iteration is linear with the number of initial hyperplane arrangements.
This process provides us a set of arrangements with low distortion

By repeating the process multiple times, we can accordingly generate a set of good initialization points. 

Pseudo-code of the proposed genetic algorithm is also described in Algo. \ref{al:genetic}.

\begin{algorithm}
\caption{Genetic algorithm}
\begin{spacing}{1.5} 
\begin{algorithmic}[1]
\STATE {\bf Input:} ${T_{\max}}$, $M$ and $k < M \in \Nbb$, a set $\{\Acal_{0,m}\}_{m \in [M]}$ of random hyperplane arrangements,  a fixed parameter
\STATE {\bf Output:} one hyperplane arrangement with low distortion
\FOR{$t$ in $[T_{\max}]$ }
\STATE order $\{\Acal_{t,m}\}_{m \in [M]}$ by decreasing fitness
\STATE $m \leftarrow k$
\STATE form pairs $(\Acal_{t,m_1} , \Acal_{t,m_2})_{i}$ (read below about the 2 existing policies to chose $(m_1,m_2)$ pairs)
\FOR {each pair $(m_1,m_2)$}
\STATE perform crossover between $\Acal_{t,m_1}$ and $\Acal_{t,m_2}$
\ENDFOR 
\STATE order $\{\Acal_{t,m}\}_{m \in [M]}$ by decreasing fitness
\FOR{$m \in [M]$}
\IF{$m < \frac M 2$}
\STATE $\Acal_{t+1,m} \leftarrow \Acal_{t,m}$
\ELSE
\STATE $\Acal_{t+1,m} \leftarrow mutate(\Acal_{t,m})$
\ENDIF
\ENDFOR
\ENDFOR
\STATE select ${\max(\{\Acal_{T_{\max},m}\}}_{m \in [M]})$
\end{algorithmic}
\end{spacing}
\label{al:genetic}
\end{algorithm}

\begin{Remark} \color{black}
An idea that naturally arises would be to consider all possible hyperplane configurations and simply optimize each of them. In App. \ref{sec:Some notion of matroid theory}, we explain some notions of matroid theory that provide a framework for the exhaustive generation of hyperplane configurations. In App. \ref{graph}, we define a way to describe the structure of hyperplane arrangements with graphs. Then in App. \ref{sec:Progressive arrangement growth}, we expose how one can progressively add nodes to this graph representation in order to generate graph configurations, and the reasons for this approach not being applicable in practice. 
\end{Remark}

\subsection{Optimization step}
\label{sec:Optimization step}

Being inspired from LBG algorithm (described in \ref{al:quantization}), our  quantizer design algorithm also features two steps that are being iterated until convergence:
(i) the optimization  of the set of reconstruction points {\footnotesize $\Xhv$} for a given combiner configuration $[\Vv \ \tv]$ and (ii) the optimization of the combiner configuration for a given set of reconstruction points.
Step (i) consists in choosing the reconstruction points as the centroids (with regards to the source distribution) of the regions defined by the hyperplane arrangement $[\Vv \ \tv]$ in the space $\Rbb^d$ \cite{orlik2013arrangements}.
The optimization step (ii) presents more complexity. 
In that step we update the hyperplane arrangement $[\Vv \ \tv]$ using two approaches:
a \emph{global} configuration update and a \emph{local} one as described in Algorithm \ref{al:one}. 
The global configuration update consists in adding a random perturbation to every hyperplane, perturbation which variance decreases over iterations.
With the local configuration update, one hyperplane is randomly selected among the hyperplanes of the configuration. Its position and orientation are then set so as to minimize the distortion of the resulting output.
In order to do that, we calculate the resulting MSE for several values of the hyperplane coefficients and apply interpolation between these values to find the best ones.
At each iteration, one of the two configuration update approaches is chosen at random. The probability of choosing the local configuration update augments with the number of iterations, at a speed depending of an empirically fixed parameter $s$. 

\begin{figure}
	\centering
	\includegraphics[width=1\linewidth]{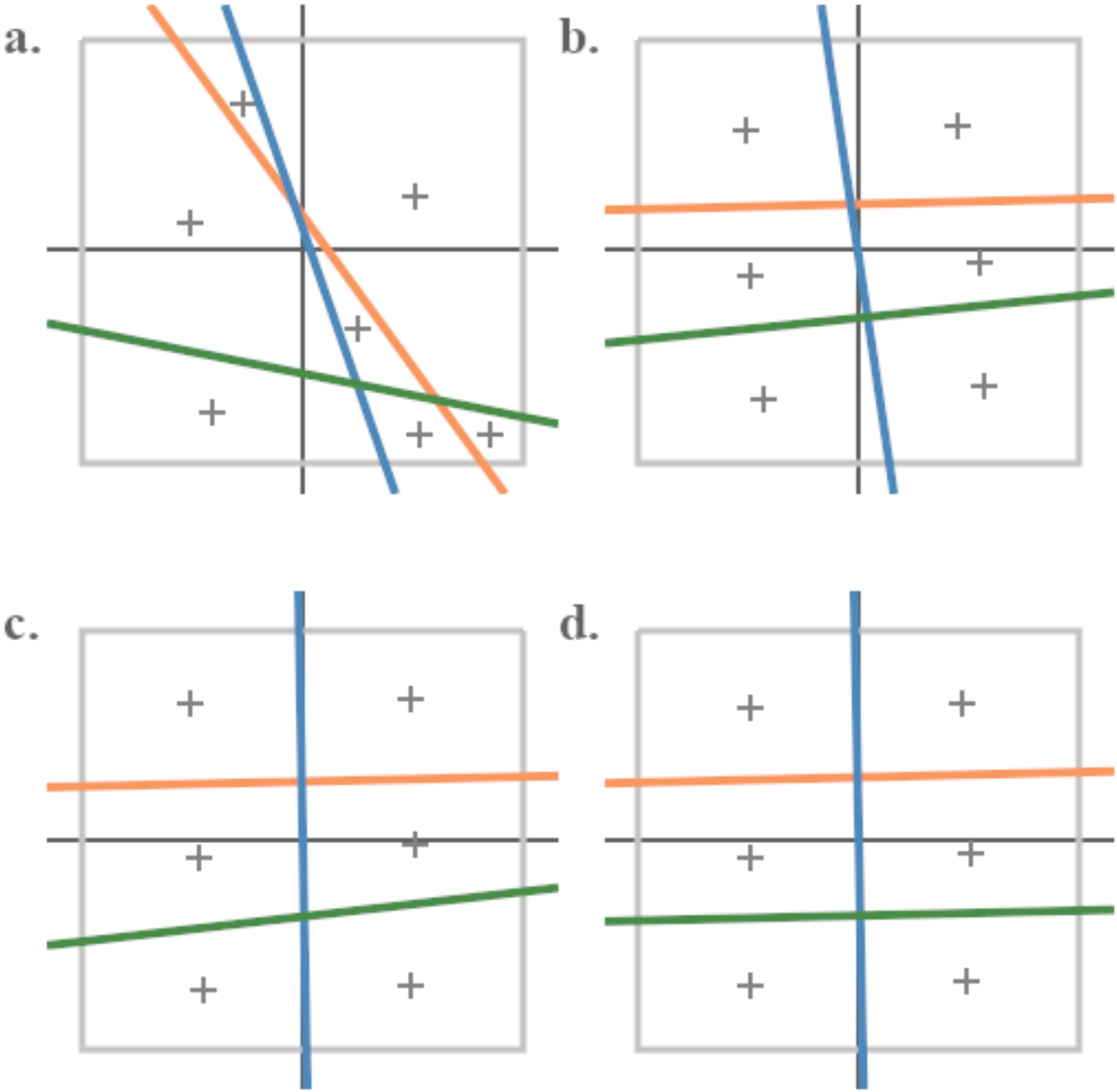}
	\caption{
	Hyperplane arrangement at different steps of the optimization algorithm. The source distribution is unitary uniform and its support is represented by the grey square. The crosses show the centroids of each region. The regions are becoming more and more regular over iterations.
    }
	\label{fig: optimisation-steps-uniform3}
\end{figure}

The reasoning behind differentiating local and global update is the following.
Consider a lower triangular matrix $M$ of size $\rr(d,\ntq) \times \rr(d,\ntq)$ and let the element in position  $i\times j$ in $M$ be equal to one of the hyperplanes separating $\Xh_i$ and $\Xh_j$ if such reconstruction point exists or equal to zero otherwise. \footnote{Note that some hyperplane arrangements induce less that $\rr(d,\ntq)$ regions: we assume that there exists a natural numbering of the possible $\rr(d,\ntq)$ regions.} The matrix $M$ can be thought of as one among a finite number of ways in which hyperplanes separate the reconstruction points.

In this view, the local update maximizes the quantizer performance in a given value of $M$. The global update, instead, allows ``hyperplanes to jump over centroids'', resulting in a different  matrix $M$. 
This is illustrated in Fig. \ref{fig: optimisation-steps-uniform3}, where passing from step a. to step b. requires a global update (that is changing the structure of the arrangement) but passing from step b. to c. and c. to d. only require local updates (optimizing the configuration).

\begin{algorithm}
\caption{Quantizer design algorithm}
\begin{spacing}{1.5} 
\begin{algorithmic}
\STATE {\bf Input:} $s$ and $[T_{\max}] \in \Nbb$, a hyperplane arrangement $\Acal_0$, $d$ the dimension
\STATE {\bf Output:} a hyperplane arrangement for quantization
\FOR{$t$ in $[T_{\max}]$}
\IF{$random(0,1) < \exp(-s.t)$}
\STATE let $\Hcal_{i,t}$ be the $i^{th}$ hyperplane of $\Acal_t$ 
\STATE let $\Vcal_{i,t}$ be a vector of $d$ realisations of $\Ncal(0,\sigma_t)$ (with $\sigma_t$ a parameter decreasing over $t$)
\STATE for all $i$, $\Hcal_{i,t} \leftarrow \Hcal_{i,t} + \Vcal_{i,t}$ 
\ELSE
\STATE select one hyperplane $\Hcal$ of $\Acal_t$
\FOR{$i$ in $[d]$}
\STATE let $p_i$ be the $i^{th}$ coefficient of $\Hcal$ and  compute distortion for alternate values of $p_i$
\STATE interpolate the function $f:p_i\mapsto distortion$ 
\STATE $p_i \leftarrow \min(f)_{p_i \in \Rbb}$
\ENDFOR
\ENDIF
\ENDFOR
\end{algorithmic}
\end{spacing}
\label{al:one}
\end{algorithm}

\subsection{Discussion}

A fundamental element in step (ii) of the design algorithm is the evaluation of the distortion (the MSE in our case) for a given hyperplane arrangement and set of reconstruction points. 
Given numerical precision limitations, our MSE evaluation has to be an approximation method, that uses numerical integration methods and particle filters.
{\color{black}
More precisely, we generate random points following the source distribution,  until a certain number of points is reached for every region of the hyperplane arrangement.  
For each random point generated, the output of the quantizer is the reconstruction point of the region in which that random point is located.
Finally, the average of the squared distance between the random points and their corresponding output point is taken as the estimation of the MSE.
}

We use a similar approach to estimate the centroids of the regions formed by a hyperplane arrangement, since that estimation requires numerical integration too.
To estimate the centroid positions, random points are also generated following the source distribution. Each point is assigned to the region it is in. We can then obtain the position of each centroid by computing the average coordinates of all the points generated in each region.

{\color{black} Note that these two steps (estimation of the centroids and estimation of the MSE) are done separately. In particular, they use distinct samples of the input distribution. }

With the alternance of steps (i) and (ii) of the algorithm, the hyperplane configuration converges to an optimum.
Multiple random restarts of the algorithm occasionally lead to convergence to distinct local minima.
%
%
These minimal values are due either to a limitation in the precision of the  numerical integration or to a local minimum in the quantizer configuration. 
In the next section, we shall further comment on this and other aspects of the proposed optimization algorithm in view of numerical evaluation results.

\section{Numerical evaluations}
\label{sec:Simulation results}

In this section, we numerically investigate the a few aspects of the proposed algorithm. 
In  Sec. \ref{sec:Genetic Initialization Performance}   we will investigate the performance of the genetic algorithm discussed in Sec.  \ref{sec:genetic initialization}.
In Sec.  \ref{sec:Optimization Step Performance}, we discuss the optimization step in Sec. \ref{sec:Optimization step}. 
Finally, in Sec. \ref{sec:LBG performance comparison}, we provide a performance comparison of the proposed algorithm  and the classic LBG algorithm in Sec. \ref{sec:Quantization Algorithm}. 

\subsection{Genetic Initialization Performance}
\label{sec:Genetic Initialization Performance} 

Let us return to the genetic initialization discussed in Sec.  \ref{sec:genetic initialization}. 
The nature of this genetic algorithm is such that the performance depends on the number of comparator configurations in input to the algorithm (see line 1 of Algo \ref{al:genetic}). 
The more  are the available configurations, the bigger is the pool of genes that can be used.
On the other hand,  adding too many configurations can be computationally expensive.
%

\begin{figure}[H]
    \centering
    \begin{subfigure}[t]{0.475\linewidth}
    	\centering
    	\begin{tikzpicture}
\definecolor{mycolor1}{rgb}{0.5,0.5,0.8}%
\definecolor{mycolor2}{rgb}{0.00000,0.44706,0.74118}%
\definecolor{mycolor3}{rgb}{0.00000,0.0,0.49804}%
\definecolor{mycolor4}{rgb}{0.87059,0.49020,0.00000}%
\definecolor{mycolor5}{rgb}{0.00000,0.44700,0.74100}%
\definecolor{mycolor6}{rgb}{0.74902,0.00000,0.74902}%

\begin{axis}[%
font=\small,
width=\textwidth,
height=0.7\textwidth,
scale only axis,
width=.8\linewidth,
xmin=-1,
xmax=30,
xlabel style={font=\small\color{white!15!black},yshift=.2cm},
xlabel={iterations},
ymin=0.35,
ymax=0.5,
ylabel style={font=\small\color{white!15!black},yshift=-.3cm},
ylabel={distortion},
axis background/.style={fill=white},
xmajorgrids,
ymajorgrids,
legend style={nodes={scale=0.85, transform shape}, legend cell align=left, align=left, draw=white!15!black}
]

\addplot [color=mycolor3, line width=1.5pt, mark=dot, mark options={solid, mycolor3}]
  table[row sep=crcr]{%
0   0.4603304228689396 \\
1   0.44993026589490254 \\
2   0.4474148273943441 \\
3   0.44603960208864746 \\
4   0.4454072711066659 \\
5   0.4314205641399977 \\
6   0.43294707519221753 \\
7   0.4316500616066886 \\
8   0.4174695601933618 \\
9   0.409404818903425 \\
10   0.41053049278354503 \\
11   0.41032149045511296 \\
12   0.4085127509176299 \\
13   0.40468235868555313 \\
14   0.40115969990771033 \\
15   0.3967866922627108 \\
16   0.39693362062402976 \\
17   0.39754823200683276 \\
18   0.39716767785820495 \\
19   0.3857266407111044 \\
20   0.3856821896746899 \\
21   0.38555814598428617 \\
22   0.38035409238285767 \\
23   0.3798135323461288 \\
24   0.37996319805183987 \\
25   0.3796198644033937 \\
26   0.3792743928708372 \\
27   0.37846649452657344 \\
28   0.3774897518647482 \\
29   0.3769420780160302 \\
};
\addlegendentry{average of several executions}

\addplot [color=mycolor1, line width=1.5pt, mark=dot, mark options={solid, mycolor1}]
  table[row sep=crcr]{%
0   0.4513257176888891 \\
1   0.43133293804100226 \\
2   0.4261324242890473 \\
3   0.4268321709453043 \\
4   0.42542908748741987 \\
5   0.42545457659434316 \\
6   0.42587785342780565 \\
7   0.4249404383344752 \\
8   0.3954805489831661 \\
9   0.39431772971157725 \\
10   0.3948566765027272 \\
11   0.39499103972812355 \\
12   0.39514931471789233 \\
13   0.3878530324117235 \\
14   0.3882777290737547 \\
15   0.38788349858102084 \\
16   0.38763120625155684 \\
17   0.3882497240176939 \\
18   0.3873492281589142 \\
19   0.3742710236017977 \\
20   0.3735902556056926 \\
21   0.3742670549863208 \\
22   0.3742789706618358 \\
23   0.3743224832927469 \\
24   0.3751052090284273 \\
25   0.37422733699006844 \\
26   0.3740516684417704 \\
27   0.3744365929901869 \\
28   0.37387247084198894 \\
29   0.37425844617858406 \\
};

\addplot [color=mycolor1, line width=1.5pt, mark=dot, mark options={solid, mycolor1}]
  table[row sep=crcr]{%
0   0.4693351280489901 \\
1   0.4685275937488028 \\
2   0.46869723049964085 \\
3   0.4652470332319906 \\
4   0.46538545472591186 \\
5   0.4373865516856522 \\
6   0.44001629695662947 \\
7   0.43835968487890203 \\
8   0.43945857140355754 \\
9   0.42449190809527276 \\
10   0.42620430906436285 \\
11   0.42565194118210237 \\
12   0.42187618711736746 \\
13   0.42151168495938274 \\
14   0.414041670741666 \\
15   0.4056898859444008 \\
16   0.4062360349965027 \\
17   0.4068467399959716 \\
18   0.40698612755749564 \\
19   0.39718225782041106 \\
20   0.39777412374368726 \\
21   0.3968492369822516 \\
22   0.38642921410387954 \\
23   0.3853045813995106 \\
24   0.3848211870752524 \\
25   0.385012391816719 \\
26   0.38449711729990393 \\
27   0.38249639606296004 \\
28   0.3811070328875074 \\
29   0.3796257098534763 \\
};
\addlegendentry{distinct executions of the algorithm}

\end{axis}
\end{tikzpicture}%
    	\caption{Two random restarts of the genetic algorithm in gray and their average in blue, on 30 iterations, 3 hyperplanes in 3 dimensions.	}
    	\label{fig:avg3d3h}
    \end{subfigure}
    \hfill
    \begin{subfigure}[t]{0.475\linewidth}
    	\centering
    	\begin{tikzpicture}
\definecolor{mycolor1}{rgb}{0.5,0.5,0.8}%
\definecolor{mycolor2}{rgb}{0.00000,0.44706,0.74118}%
\definecolor{mycolor3}{rgb}{0.00000,0.0,0.49804}%
\definecolor{mycolor4}{rgb}{0.87059,0.49020,0.00000}%
\definecolor{mycolor5}{rgb}{0.00000,0.44700,0.74100}%
\definecolor{mycolor6}{rgb}{0.74902,0.00000,0.74902}%

\begin{axis}[%
font=\small,
width=\textwidth,
height=0.7\textwidth,
scale only axis,
width=.8\linewidth,
xmin=-1,
xmax=30,
xlabel style={font=\small\color{white!15!black},yshift=.2cm},
xlabel={iterations},
ymin=0.3,
ymax=0.4,
ylabel style={font=\small\color{white!15!black},yshift=-.3cm},
ylabel={distortion},
axis background/.style={fill=white},
xmajorgrids,
ymajorgrids,
legend style={nodes={scale=0.85, transform shape}, legend cell align=left, align=left, draw=white!15!black}
]

\addplot [color=mycolor3, line width=1.5pt, mark=dot, mark options={solid, mycolor3}]
  table[row sep=crcr]{%
0   0.380171739703848 \\
1   0.372731757312579 \\
2   0.3698514898725891 \\
3   0.371766285086517 \\
4   0.3559783989564015 \\
5   0.35699463807633863 \\
6   0.35575864099453636 \\
7   0.35099426071309003 \\
8   0.35042993181635906 \\
9   0.35137964325863424 \\
10   0.351383422791553 \\
11   0.35036150363559926 \\
12   0.33089906857069556 \\
13   0.32624026926841193 \\
14   0.32662972245126654 \\
15   0.3261829696104023 \\
16   0.32624576679982586 \\
17   0.3266544334011223 \\
18   0.32670901682695697 \\
19   0.32652922406140716 \\
20   0.326146708677588 \\
21   0.3259225840187412 \\
22   0.32616004149286726 \\
23   0.3265750143549162 \\
24   0.3261641714984617 \\
25   0.32603814595593805 \\
26   0.3263844228294495 \\
27   0.3258994932476757 \\
28   0.32574176185473486 \\
29   0.32633835840943465 \\
};
\addlegendentry{average of several executions}

\addplot [color=mycolor1, line width=1.5pt, mark=dot, mark options={solid, mycolor1}]
  table[row sep=crcr]{%
0   0.36564728910590516 \\
1   0.3690205776080862 \\
2   0.3665982594861531 \\
3   0.3693959163438128 \\
4   0.3506937886752435 \\
5   0.34952824079947187 \\
6   0.35035797964877824 \\
7   0.34987415369679314 \\
8   0.3492902886788179 \\
9   0.35047977576196704 \\
10   0.3498643883659069 \\
11   0.348801732545868 \\
12   0.31950956948567916 \\
13   0.31954211201452354 \\
14   0.3199259252018874 \\
15   0.31906404729898236 \\
16   0.31917016879335136 \\
17   0.31973251567562583 \\
18   0.31961157046583133 \\
19   0.3196196829644505 \\
20   0.3192402701541002 \\
21   0.3185892977694092 \\
22   0.31924057441715753 \\
23   0.31943421332433924 \\
24   0.3188677741652305 \\
25   0.31885598549062166 \\
26   0.3196058329357453 \\
27   0.3189600045911284 \\
28   0.3190497261558497 \\
29   0.319526114924891 \\
};

\addplot [color=mycolor1, line width=1.5pt, mark=dot, mark options={solid, mycolor1}]
  table[row sep=crcr]{%
0   0.3946961903017909 \\
1   0.3764429370170718 \\
2   0.3731047202590251 \\
3   0.3741366538292212 \\
4   0.3612630092375595 \\
5   0.3644610353532054 \\
6   0.3611593023402945 \\
7   0.352114367729387 \\
8   0.3515695749539002 \\
9   0.3522795107553015 \\
10   0.35290245721719904 \\
11   0.35192127472533047 \\
12   0.342288567655712 \\
13   0.3329384265223003 \\
14   0.33333351970064573 \\
15   0.33330189192182214 \\
16   0.33332136480630037 \\
17   0.33357635112661876 \\
18   0.33380646318808266 \\
19   0.3334387651583639 \\
20   0.33305314720107576 \\
21   0.33325587026807324 \\
22   0.33307950856857704 \\
23   0.3337158153854932 \\
24   0.33346056883169295 \\
25   0.3332203064212544 \\
26   0.33316301272315374 \\
27   0.332838981904223 \\
28   0.33243379755362 \\
29   0.33315060189397827 \\
};
\addlegendentry{distinct executions of the algorithm}

\end{axis}
\end{tikzpicture}%
    	\caption{Two random restarts of the genetic algorithm in gray and their average in blue on 30 iterations, 4 hyperplanes in 3 dimensions.}
    	\label{fig:avg3d4h}
    \end{subfigure}
    \vspace{0.5cm}
    \caption{Genetic algorithm evolution.}
    \label{Fig: genetic_evolution}
\end{figure}
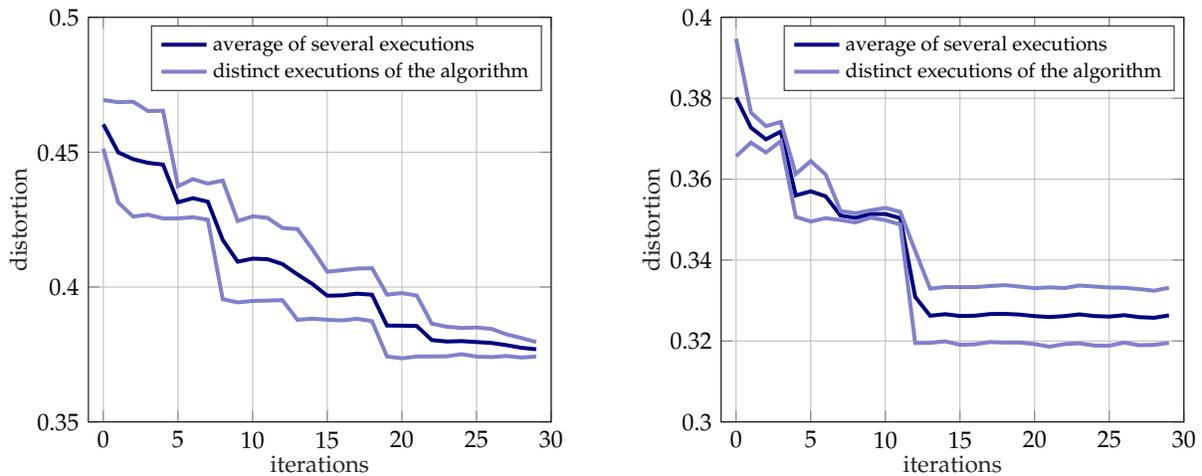

{\color{black} 
Fig. \ref{Fig: genetic_evolution} shows the result obtained by the genetic algorithm with an input of dimension 3 and configurations of 3 or 4 hyperplanes in Fig. \ref{fig:avg3d3h} and Fig. \ref{fig:avg3d4h} respectively. Every run of the algorithm used a pool of 10 configurations. These configurations are then mixed following the steps described in Sec. \ref{sec:genetic initialization}. The algorithm was run twice for \ref{fig:avg3d3h} and twice again for \ref{fig:avg3d4h}. 
Both (a) and (b) display these two runs of the algorithm in light blue, and their average in dark blue.
}

Note that in these two particular executions, in both Fig. \ref{fig:avg3d3h} and Fig. \ref{fig:avg3d4h}, that the distortion decreases very irregularly.
This curve is composed of flat sections and a few sudden drops: this suggests that improvements happen from time to time when the right combination of two configurations is formed. 
For this reason, added to the general propensity of genetic algorithms to converge to local minima, this algorithm cannot be used for the whole optimization process. 
However, its computational efficiency makes it a good pre-optimization tool. Indeed, the distortion function is the often the most computationally heavy part of the optimization. This implementation calls to the distortion function $2\times n$ times per iteration (with n the number of configurations), while the previous algorithm we described in Sec. \ref{sec:Design Algorithm} calls it $40\times H\times d$ with $H$ the number of hyperplanes and $d$ the dimension.
Thus, this genetic algorithm can be used to quickly obtain a rough estimate of the global minimum, before a more precise and more computation-costly algorithm can finish the optimization.

\subsection{Optimization Step Performance}
\label{sec:Optimization Step Performance}

\begin{figure}
	\centering
	\includegraphics[width=1\linewidth]{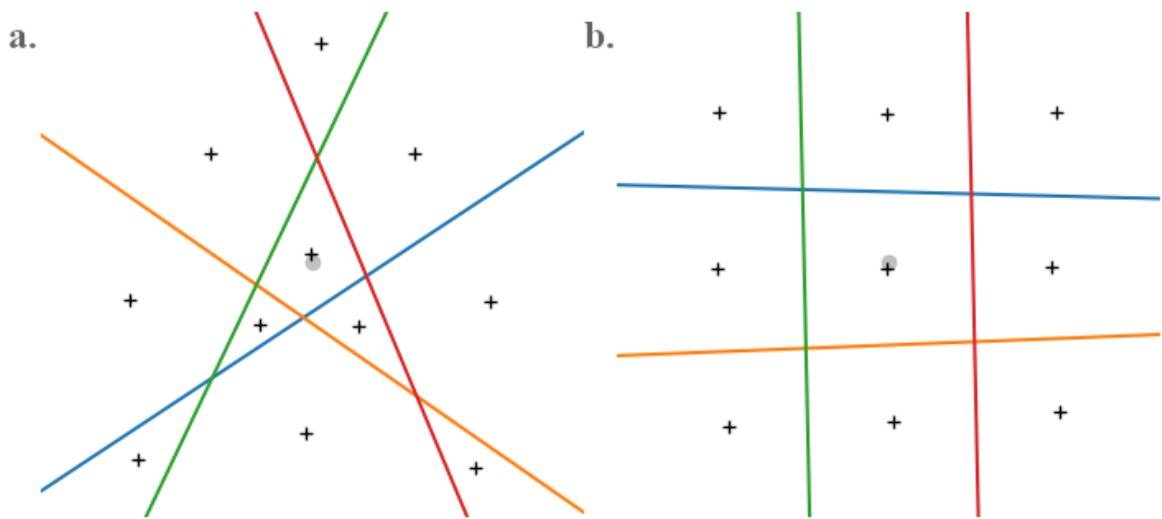}
	\caption{2 local minima obtained with the algorithm in Sec. \ref{sec:Design Algorithm}. The center of the Gaussian source distribution is marked by the grey dot, and the centroid of each region is indicated by a cross.}
	\label{fig:local-minima}
\end{figure}

Next, we numerically investigate the optimization step discussed in Sec. \ref{sec:Optimization step}. As discussed in this section, the proposed alternate optimization encounters a number of local minima:
Fig. \ref{fig:local-minima} (showing the quantizers obtained by 2 distinct random restarts of the optimization algorithm) demonstrates the possibility of convergence to distinct local minima.%
While the arrangement in Fig. \ref{fig:local-minima}.a  has 6 centroids, the one in Fig \ref{fig:local-minima}.b  has 7, though both of them score similar performance for a standard Gaussian source distribution. 
Regarding the structure of the arrangement, however, these two configurations are rather different, and the proposed algorithm is unable to go from the configuration in Fig.\ref{fig:local-minima}.a  to the slightly better configuration \ref{fig:local-minima}.b.



\subsection{LBG performance comparison}
\label{sec:LBG performance comparison}

Finally, we come to perhaps the  most interesting numerical evaluations: the comparison between the proposed quantization approach and the classic LBG algorithm. 
Such comparison indeed addresses the observation in Fig. \ref{fig:quantization} which is the initial motivation behind the proposed approach. 

This section presents the performance of the quantizer described in Sec. \ref{sec:Vector Quantizer Model} with a configuration resulting from the algorithm in  Sec. \ref{sec:Design Algorithm} for the case of %
(i)  standard Gaussian and (ii) unitary uniform distribution. We obtained similar performance results in both cases.
In all cases, we compared the performance attained with the performance of the quantizer obtained by the classic LBG algorithm.

\begin{figure}[t]
	\centering
    \begin{tikzpicture}
\definecolor{mycolor1}{rgb}{0.63529,0.07843,0.18431}%
\definecolor{mycolor2}{rgb}{0.00000,0.44706,0.74118}%
\definecolor{mycolor3}{rgb}{0.00000,0.49804,0.00000}%
\definecolor{mycolor4}{rgb}{0.87059,0.49020,0.00000}%
\definecolor{mycolor5}{rgb}{0.00000,0.44700,0.74100}%
\definecolor{mycolor6}{rgb}{0.74902,0.00000,0.74902}%

\begin{axis}[%
font=\small,
width=13cm,
height=7cm,
scale only axis,
xmin=0.9,
xmax=3.6,
xlabel style={font=\color{white!15!black}},
xlabel={Number of hyperplanes or regions, divided by dimension},
ymin=0,
ymax=0.7,
ylabel style={font=\color{white!15!black}},
ylabel={Distortion},
axis background/.style={fill=white},
xmajorgrids,
ymajorgrids,
legend style={nodes={scale=0.85, transform shape}, legend cell align=left, align=left, draw=white!15!black}
]
\addplot [color=mycolor1, only marks, line width=1.5pt, mark=asterisk, mark options={solid, mycolor1}]
  table[row sep=crcr]{%
1.0   0.6816569059729831 \\
1.5   0.4630301264439741 \\
2.0   0.36386429963027117 \\
2.5   0.310145281431466 \\
3.0   0.2614815246789771 \\
3.5   0.22933794535213367 \\
1.0   0.6422926694139711 \\
1.3333333333333333   0.5298363873098486 \\
1.6666666666666667   0.4706963995974392 \\
2.0   0.4162606529077224 \\
2.3333333333333335   0.38582421439866893 \\
1.0   0.6471639433962179 \\
1.25   0.5778122228249278 \\
1.5   0.5386808350480564 \\
1.75   0.5020466111669821 \\
1.0   0.6622103624707472 \\
1.2   0.6179865744713496 \\
1.4   0.5868711103370133 \\
};
\addlegendentry{LBG algorithm}

\addplot [color=mycolor2, only marks, line width=1.5pt, mark=o, mark options={solid, mycolor2}]
  table[row sep=crcr]{%
1.0   0.36518307188954274 \\
1.5   0.2842680008074216 \\
2.0   0.20973150242889144 \\
2.5   0.17342372895962915 \\
3.0   0.14364803102650017 \\
3.5   0.12440754634447539 \\
1.0   0.36882947775408437 \\
1.3333333333333333   0.30747127729838925 \\
1.6666666666666667   0.2607921225139601 \\
2.0   0.2278838570833841 \\
2.3333333333333335   0.20750843162915586 \\
1.0   0.373279821597409 \\
1.25   0.33196443108803386 \\
1.5   0.2958362754840413 \\
1.75   0.2742152504958007 \\
1.0   0.3756908929084427 \\
1.2   0.36196649725327357 \\
1.4   0.3279794047938959 \\
};
\addlegendentry{proposed algorithm}

\end{axis}
\end{tikzpicture}%
	\caption{Diagram showing the average distortion (Mean Squared-Error here) values attained by the classic LBG algorithm (green) and the proposed algorithm (blue) {\color{black}on a Gaussian input signal. The number $d$ of dimensions ranges from 2 to 6, and the number of regions (for LBG) or hyperplanes (for the proposed algorithm) ranges from $d$ to 8. The results are averaged over several random restarts of the algorithms}. For better readability, these results are spread on a X-axis representing the ratio between number of hyperplanes or regions, and dimension.}  \label{fig:general-comparison}
\end{figure}
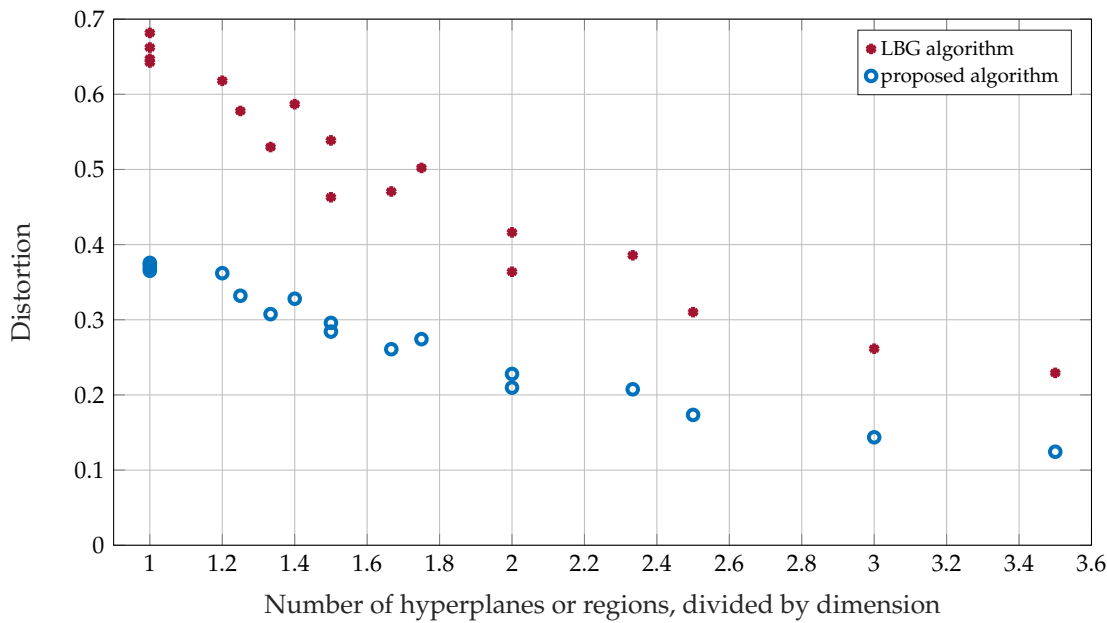

\smallskip
The code for this algorithm is available online: see \cite{coderepo}.

\subsubsection{Gaussian distribution}

\begin{figure}
    \centering
    \begin{subfigure}[t]{0.8\linewidth}
    	\centering
    	\begin{tikzpicture}
\definecolor{mycolor1}{rgb}{0.73529,0.07843,0.18431}%
\definecolor{mycolor2}{rgb}{0.00000,0.44706,0.74118}%
\definecolor{mycolor3}{rgb}{0.2,0.6,0.9}%
\definecolor{mycolor4}{rgb}{0.87059,0.49020,0.00000}%
\definecolor{mycolor5}{rgb}{0.00000,0.44700,0.74100}%
\definecolor{mycolor6}{rgb}{0.74902,0.00000,0.74902}%

\begin{axis}[%
font=\small,
height=0.5\textwidth,
scale only axis,
width=.9\linewidth,
xmin=0.9,
xmax=5.1,
xlabel style={font=\small\color{white!15!black},yshift=.2cm},
xlabel={Number of comparators},
ymin=0,
ymax=1.6,
ylabel style={font=\small\color{white!15!black},yshift=-.4cm},
ylabel={Distortion},
axis background/.style={fill=white},
xmajorgrids,
ymajorgrids,
legend style={nodes={scale=0.85, transform shape}, legend cell align=left, align=left, draw=white!15!black}
]

\addplot [color=mycolor3, line width=1.5pt, mark=o, mark options={solid, mycolor3}]
  table[row sep=crcr]{%
1 1.37\\
2 0.72\\
3 0.445\\
4 0.304\\
5 0.219\\
};
\addlegendentry{LBG optimal quantizer (for comparison)}

\addplot [color=mycolor2, line width=1.5pt, mark=o, mark options={solid, mycolor2}]
  table[row sep=crcr]{%
1 1.37\\
3 0.926\\
5 0.72\\
};
\addlegendentry{LBG, same number of comparators}

\addplot [color=mycolor1, line width=1.5pt, mark=o, mark options={solid, mycolor1}]
  table[row sep=crcr]{%
1 1.37\\
2 0.728\\
3 0.565\\
4 0.420\\
5 0.344\\
};
\addlegendentry{proposed algorithm}

\end{axis}
\end{tikzpicture}%
    	\caption{Quantizer performance for a Gaussian input distribution in $\Rbb^2$.}
    	\label{fig: comparison-g}
    \end{subfigure}
    
    \vspace{1cm}
    
    \begin{subfigure}[t]{0.8\linewidth}
    	\centering
    	\begin{tikzpicture}
\definecolor{mycolor1}{rgb}{0.73529,0.07843,0.18431}%
\definecolor{mycolor2}{rgb}{0.00000,0.44706,0.74118}%
\definecolor{mycolor3}{rgb}{0.2,0.6,0.9}%
\definecolor{mycolor4}{rgb}{0.87059,0.49020,0.00000}%
\definecolor{mycolor5}{rgb}{0.00000,0.44700,0.74100}%
\definecolor{mycolor6}{rgb}{0.74902,0.00000,0.74902}%

\begin{axis}[%
font=\small,
height=0.5\textwidth,
scale only axis,
width=.9\linewidth,
xmin=0.9,
xmax=5.1,
xlabel style={font=\small\color{white!15!black},yshift=.2cm},
xlabel={Number of comparators},
ymin=0,
ymax=0.5,
ylabel style={font=\small\color{white!15!black},yshift=-.4cm},
ylabel={Distortion},
axis background/.style={fill=white},
xmajorgrids,
ymajorgrids,
legend style={nodes={scale=0.85, transform shape}, legend cell align=left, align=left, draw=white!15!black}
]

\addplot [color=mycolor3, line width=1.5pt, mark=o, mark options={solid, mycolor3}]
  table[row sep=crcr]{%
1 0.42\\
2 0.16\\
3 0.105\\
4 0.068\\
5 0.045\\
};
\addlegendentry{LBG optimal quantizer (for comparison)}

\addplot [color=mycolor2, line width=1.5pt, mark=o, mark options={solid, mycolor2}]
  table[row sep=crcr]{%
1 0.42\\
3 0.26\\
5 0.17\\
};
\addlegendentry{LBG, same number of comparators}

\addplot [color=mycolor1, line width=1.5pt, mark=o, mark options={solid, mycolor1}]
  table[row sep=crcr]{%
1 0.42\\
2 0.165\\
3 0.12\\
4 0.081\\
5 0.062\\
};
\addlegendentry{proposed algorithm}

\end{axis}
\end{tikzpicture}%
    	\caption{Quantizer performance for a uniform unitarian input distribution in $\Rbb^2$.}
    	\label{fig: comparison-u}
    \end{subfigure}
    \vspace{0.5cm}
    \caption{
    The red curve shows the distortion of the proposed algorithm and the dark blue curve shows the distortion from the quantizer of LBG algorithm, in function of the number of comparators available.
    For comparison purpose, the light blue curve shows the performance of a quantizer obtained with LBG algorithm, but with as many regions as the quantizer obtained with the proposed algorithm (using more comparators than the proposed algorithm).
    For each plot, the values were averaged over numerous restarts of the algorithm.
    }
    \label{fig:comparisons}
\end{figure}
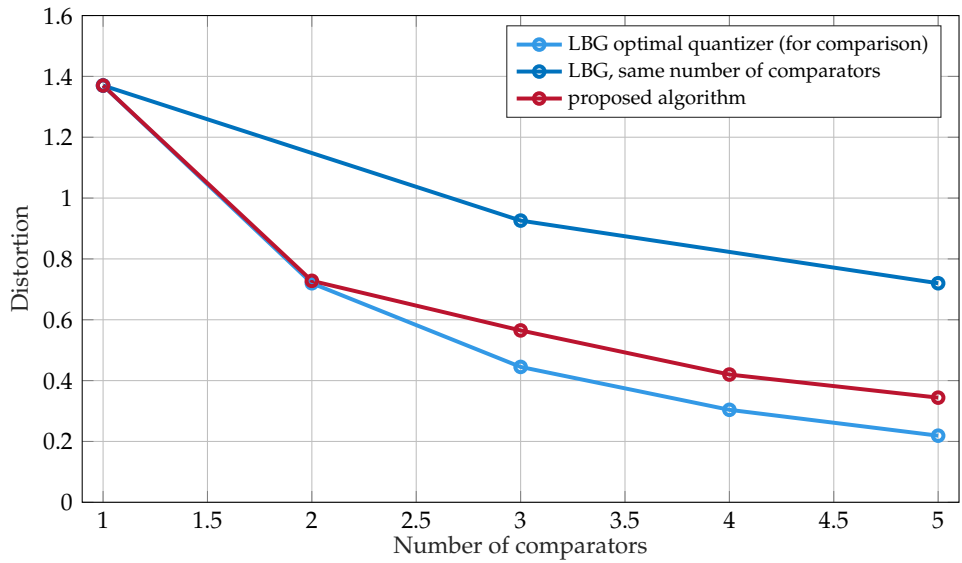
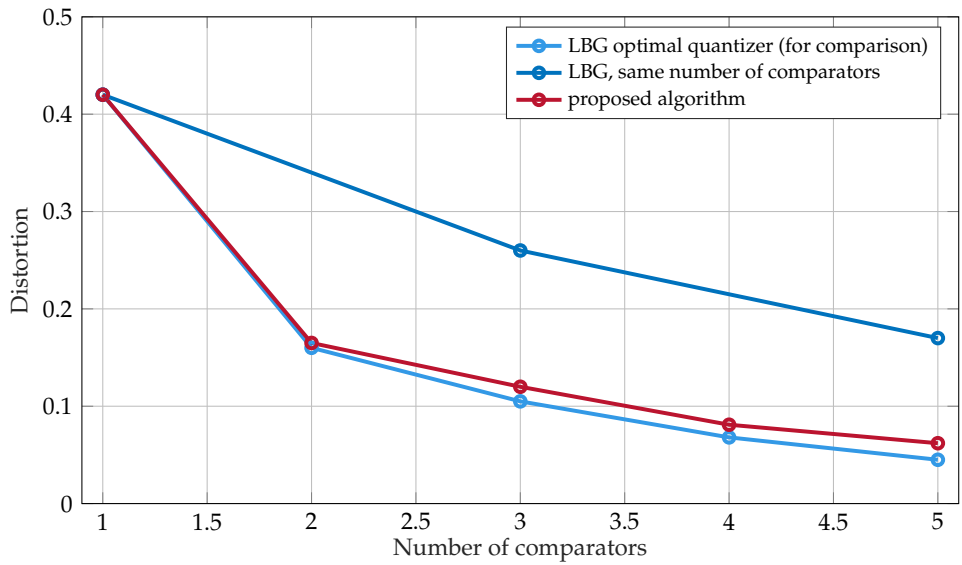

As one can observe on Fig. \ref{fig: comparison-g}, the quantizer designed by our algorithm performs better than the quantizer design using LBG algorithm. For a quantizer using 5 comparators and an input signal in $\Rbb^2$, its distortion is 0.64 that of LBG quantizer.
{\color{black}
Fig. \ref{fig:general-comparison} presents the performance of the output of LBG algorithm, compared to that of the proposed algorithm. One can observe there that the quantizer designed by the proposed algorithm performs better than the quantizer designed by LBG algorithm in every configuration represented on the figure.
}

As could be expected, the quantizer obtained with the proposed algorithm performs slightly worse than the optimal quantizer obtained with LBG algorithm using the same number of reconstruction points (ie the same output bit rate). This shows that using the CLVQ framework allows for quantizer designs that cost little degradation in performance in the classic quantization (where the constraint is on the cardinality of the output), but generate important performance gains in the view of comparison-limited quantization.

A result that may be surprising and seems counter-intuitive is that the hyperplanes do not always form as many regions as possible. Curiously, the hyperplanes sometimes converge to positions in which they form less regions than the maximum possible, but with more regular shapes, which implies a lesser probability of values at a long distance from the representation point, thus a lower MSE. Fig. \ref{fig:local-minima}.b is an example of such configuration.
These configurations turn out to perform relatively well and sometimes better than configurations with more regions.

\subsubsection{Uniform distribution}

Running the algorithm with a unitary uniform distribution as source distribution gives similar results as with a Gaussian distribution. In this case again, our algorithm performs better than LBG with the same number of comparators. For a quantizer using 5 comparators and an input signal in $\Rbb^2$, its distortion is 0.72 that of LBG quantizer. It also performs slightly worse than LBG algorithm with the same number of reconstruction points.

Because of the square shape of the distribution support, the hyperplanes are even more likely to form rectangular regions with a uniform distribution than with the Gaussian distribution, as Fig. \ref{fig: optimisation-steps-uniform3} illustrates.

\subsubsection{Entropy maximization}
\label{sec:entropy_maximization}
As commented in Sec. \ref{sec:Discussion}, one is often interested in characterizing the entropy of the quantizer output, as this indicates whether it is necessary to perform a further compression of the quantizer output. 
%
Setting as the objective as maximization of the entropy of the output, instead of minimization of the MSE, does not require one to make any significant changes to the optimization.

In our simulations, the result of the entropy maximization is rather similar to that of the MSE minimization. 
%
The most notable difference in using entropy as a measure is that in cases like Fig. \ref{fig:local-minima}, it favors configurations of type Fig. \ref{fig:local-minima}.a rather than type Fig. \ref{fig:local-minima}.b.

%
%
%
%
%
%
%
%
%
%
%
%
%
%
%
%
%

\section{Conclusion}
\label{sec:Conclusion}
In this paper, a novel paradigm for vector quantization is considered, in which the performance of the vector quantizer is constrained by the number of comparators needed to obtain the quantized signal, instead of being constrained by the output bit rate as in the classic vector quantization problem. 
%
%
We study the scenario where the vector quantizer is made of $\ntq$ comparators that each receive a linear combination of the quantizer input and a constant bias, and produce the sign of the received signal as output.
%
We focus on the task of finding the linear combinations and constant biases, given $\ntq$, the distribution of the quantizer input and a distortion measure, so as to minimize the distortion between input and output of the quantizer.

We present an algorithm to solve this optimization problem and show the performance it attains in the case of mean squared error distortion and Gaussian and uniform i.i.d. source distributions. 
%
The proposed algorithm is composed of an initialization step and a distinct optimization step.
In the initialization step, a set of possible initial configurations are produced which span the set of possible solutions.
Diverse approaches are presented for this step: in particular we propose  a genetic algorithm which can produce a rather good initial configuration for a linear complexity.
In the optimization step, the algorithm iteratively optimized the position of the reconstruction points and the comfiguration of linear combinations received by the comparators.  
We perform numerical simulations for this algorithm and compare the performance attained to that of the classic Linde-Buzo-Gray quantizer.
%
We show that one can obtain similar performance by using a smaller number of comparators than the classic approach would require.

A number of research directions remain open from this new vector quantizer architecture. In particular, we are investigating the optimal performance attainable in the limit of infinitely long vector quantizer in which the number of available comparators $\ntq$ and bits available to represent the quantizer inputs both grow to infinity at a given constant ratio $\al$. This limit should result in a rather interesting generalization of the classic distortion-rate function.

\appendix

{

\section{-- Some notion of matroid theory}
\label{sec:Some notion of matroid theory}

Oriented matroids are defined as combinatorial objects that can contain properties of linear dependence in vector spaces. Matroids have a natural way of being mapped to some objects such as sets of points or sets of hyperplanes, while retaining relations of alignment and coplanarity between their elements. %
Oriented matroids also have these properties, but add order relations to it. In consequence, convexity relations inside vector or hyperplane arrangements are also contained in oriented matroids \cite{bjorner1994}.
An oriented matroid  is defined as follows: for a linear functional $\xv^T \in \Rbb^m$, define the \emph{covector} of $\xv$
\ea{
C_{\Acal}(\xv) = \lsb \sign(\av_1^T \xv-b_1), \ldots , \sign(\av_n^T \xv-b_n)\rsb,
}  
and let us define the set of all possible covectors of $\Acal$ as
\ea{
\Lcal_{\Acal}= \lcb C_{\Acal}(\xv) \ :\ \xv \in \Rbb^m \rcb,
}
where the $\sign$ takes value zero when $\av_i^T \xv=b_i$, so that the support of $\sign$ is $\{-1,0,+1\}$.
The set $M=-([n],\Lcal_{\Acal})$ is defined as the oriented matroid associated to $\Acal$.
Using the oriented matroid framework, one is able to state questions such as the what is the set of covectors that can be generated through the arrangement of $n$  hyperplanes in dimension $m$.
%
Regrettably, Vakil showed \cite{vakil2006murphy} using  Mn\"ev's universality theorem \cite{mnev1986} that most of the stimulating problems that could be framed with oriented matroids have no computationally efficient way to be solved.

\section{-- Graph representation}
\label{graph}
In order to efficiently describe various combinations of hyperplane arrangements, we have used graphs to represent them. 
The object we need to work with has to be able to represent a hyperplane arrangement's structure rather than one particular arrangement. %
Thus, the object has to be invariant to rotation, translation and scaling and to any hyperplane modification that does not change the overall structure of the arrangement. 
%
%
%
A satisfying representation is the one shown in  Fig. \ref{fig:transformToGraph}.

\begin{figure}
	\centering
	\includegraphics[width=\linewidth]{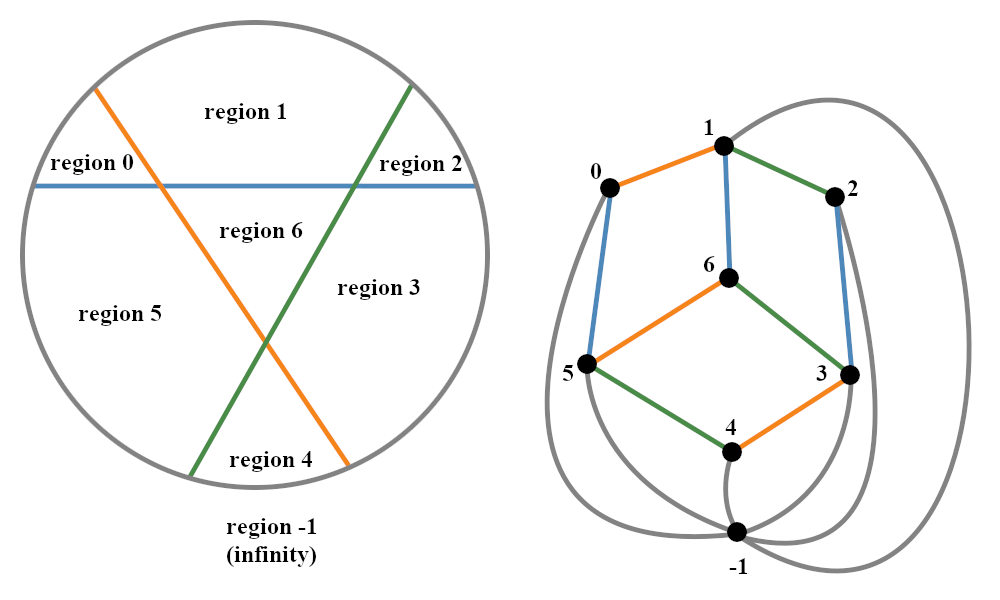}
	\caption{The left hand side is an arrangement of 3 hyperplanes that form 7 regions, with the grey circle representing infinity. The right hand side is the corresponding graph (7 vertices for the 7 regions, and one for the sphere).}
	\label{fig:transformToGraph}
\end{figure}
The interpretation for Fig. \ref{fig:transformToGraph} is the following: regions of the hyperplane arrangement (left side) are individually associated to nodes in the graph (right side).
If two regions in the arrangement share an edge, the related nodes are connected on the graph (the color of the edge on the graph corresponds to that of the hyperplane in the arrangement).
%
In consequence, the number of edges of a region is translated into the degree of the corresponding node.
%
Projectivization is used in order to accurately represent unbounded regions. A finite hypersphere is drawn around the zone of interest, and the space $\Rbb^{m}$ is projected onto that hypersphere so that coordinates at infinity are mapped to its edge.
Then, the edge of the sphere is seen as an edge of the infinite regions. This way, regions of infinite size can be fully characterized.
%
Another way to describe this representation is to define it as the dual graph of the graph that has hyperplane intersections as vertices and the hyperplanes themselves as edges. The same reasoning for using projectivization to characterize regions can be applied in a similar way.
%

Note that graph representation in Fig. \ref{fig:transformToGraph} is substantially a representation of the oriented matroid in App. \ref{sec:Some notion of matroid theory}.
Indeed, given a label for each region, these labels can be mapped to the set of all possible covectors $\Lcal_{\Acal}$ of the arrangement. 
In order to obtain such region labels, one can proceed as follows. First, give an arbitrary orientation to the hyperplanes. Each hyperplane defines two half-spaces, one of which is attributed a positive label and the other one a negative label. Then, each region can be attributed a unique label by concatenating the positive or negative labels given by each hyperplane (depending on whether the region is on the positive or negative side of the hyperplane).
Thus, each region receives a unique label composed of a number of binary variables equal to the number of hyperplanes in the arrangement.
%
Fig. \ref{fig:labelregions} shows an example of such labeling process for an arrangement of 3 hyperplanes.
\begin{figure}
	\centering
	\includegraphics[width=0.75\linewidth]{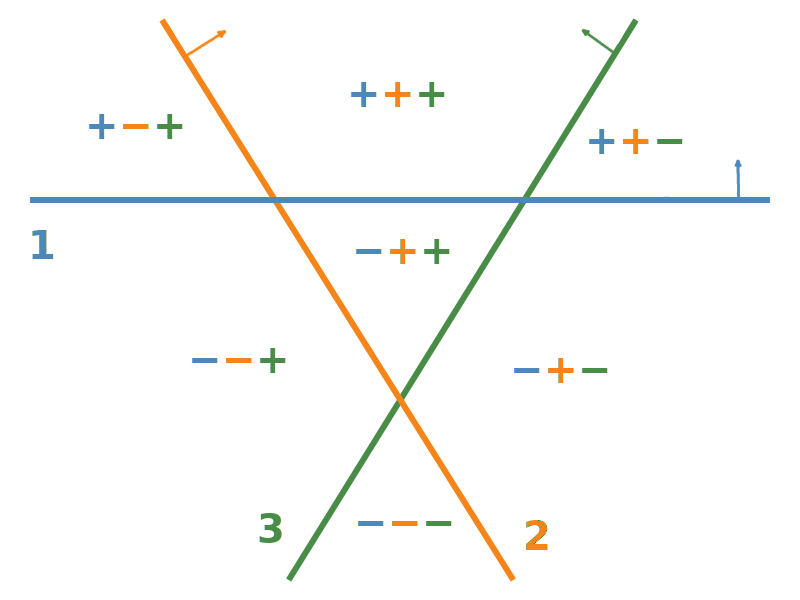}
	\caption{
	Illustration of the labeling process. In this arrangement, each of the 3 hyperplanes is oriented: it has a positive side (indicated by an arrow) and a negative side. %
	Each region is identified in a unique manner by its location relative to each hyperplane. If it is on the positive side (resp. negative side) of the $k^{th}$ hyperplane, it has a "+" (resp. "-") in the $k^{th}$ position of its label.
	}
	\label{fig:labelregions}
\end{figure}

%

\section{-- Progressive arrangement growth}
\label{sec:Progressive arrangement growth}
By numerical experimentation, it is easy to observe that the optimization step which relies on random initialization in Sec. \ref{sec:Random initialization} has a high risk of being stuck in a local minimum.
In order to alleviate this problem, we have considered a second approach to linear comparator initialization that makes use of the notions described in App. \ref{sec:Some notion of matroid theory} and of the graph representation described in App. \ref{graph}.
%
More specifically, (i) we begin by considering a simple graph with two vertices and one edge, representing two regions separated by one hyperplane then, recursively, (ii) one generates a new set of graphs by considering all  the possible ways in which a new configuration can be generated from the old one adding one more hyperplane. 
 
The recursive repetition of step (ii) above results in an exponentially growing number of configurations. Moreover, this process is susceptible to produce duplicates or equivalent configurations.
To detect and remove duplicate configurations, one can consider verifying if the graph representations of two configurations are isomorphic. 
%
%
Verifying graph isomorphisms is a well studied problem that can be solved in logarithmic time in the case of planar graphs (which is the case of representation graphs for arrangements in 2 dimensions).
However it is in sub-exponential time for non-planar graphs (and thus for higher dimension arrangements).
This presents an obstacle to the problem of determining all possible hyperplane configurations. 
More generally, there exist two other obstacles that prevent one from  generating all possible initial configurations and optimize each of them.
First,
the task of generating all these configurations is computationally very heavy. 
This is even more true in higher dimensions, as complexity grows super-exponentially with the number of dimensions. 
Second, the task of converting a hyperplane arrangement structure (that is a graph representation in our case) into an actual hyperplane arrangement with numerical values is highly complex, as is implied by Mn\"ev's universality theorem \cite{mnev1986}.
%
The high computational complexity of these two steps cause the approach not to be applicable in practice.

}



\vspace{6pt} 


\funding{The work of S. Rini is supported by the MOST grant number 109 -2927-I-009 -512.}

\acknowledgments{
We would like to give thanks to Arnau Padrol from Sorbonne Universit\'e for his help on understanding oriented matroids.
}

\conflictsofinterest{The authors declare no conflict of interest.} 

\abbreviations{The following abbreviations are used in this manuscript:\\

\noindent 
\begin{tabular}{@{}ll}
MSE & Mean Squared Error\\
LBG & Linde-Buzo-Gray algorithm\\
MIMO & Massive Input Massive Output\\
GP & General Position\\
A2D & Analog-to-Digital\\
VQ & Vector Quantization\\
CLVQ & Comparison-Limited Vector Quantization
\end{tabular}}





\reftitle{References}
\externalbibliography{yes}
\bibliography{RD_bib}

\begin{thebibliography}{-------}
\providecommand{\natexlab}[1]{#1}

\bibitem[Gray(1984)]{gray1984vector}
Gray, R.
\newblock Vector quantization.
\newblock {\em IEEE Assp Magazine} {\bf 1984}, {\em 1},~4--29.

\bibitem[Gersho and Gray(2012)]{gersho2012vector}
Gersho, A.; Gray, R.M.
\newblock {\em Vector quantization and signal compression}; Vol. 159, Springer
  Science \& Business Media,  2012.

\bibitem[Makhoul \em{et~al.}(1985)Makhoul, Roucos, and Gish]{makhoul1985vector}
Makhoul, J.; Roucos, S.; Gish, H.
\newblock Vector quantization in speech coding.
\newblock {\em Proceedings of the IEEE} {\bf 1985}, {\em 73},~1551--1588.

\bibitem[Paliwal and Atal(1993)]{paliwal1993efficient}
Paliwal, K.K.; Atal, B.S.
\newblock Efficient vector quantization of LPC parameters at 24 bits/frame.
\newblock {\em IEEE transactions on speech and audio processing} {\bf 1993},
  {\em 1},~3--14.

\bibitem[Yahampath and Rondeau(2007)]{yahampath2007multiple}
Yahampath, P.; Rondeau, P.
\newblock Multiple-description predictive-vector quantization with applications
  to low bit-rate speech coding over networks.
\newblock {\em IEEE transactions on audio, speech, and language processing}
  {\bf 2007}, {\em 15},~749--755.

\bibitem[Gersho and Ramamurthi(1982)]{gersho1982image}
Gersho, A.; Ramamurthi, B.
\newblock Image coding using vector quantization.
\newblock  ICASSP'82. IEEE International Conference on Acoustics, Speech, and
  Signal Processing. IEEE,  1982, Vol.~7, pp. 428--431.

\bibitem[Nasrabadi and King(1988)]{nasrabadi1988image}
Nasrabadi, N.M.; King, R.A.
\newblock Image coding using vector quantization: A review.
\newblock {\em IEEE Transactions on communications} {\bf 1988}, {\em
  36},~957--971.

\bibitem[Chiranjeevi and Jena(2018)]{chiranjeevi2018image}
Chiranjeevi, K.; Jena, U.R.
\newblock Image compression based on vector quantization using cuckoo search
  optimization technique.
\newblock {\em Ain Shams Engineering Journal} {\bf 2018}, {\em 9},~1417--1431.

\bibitem[Lee and Woods(1995)]{lee1995motion}
Lee, Y.Y.; Woods, J.W.
\newblock Motion vector quantization for video coding.
\newblock {\em IEEE transactions on image processing: a publication of the IEEE
  Signal Processing Society} {\bf 1995}, {\em 4},~378--382.

\bibitem[Regunathan \em{et~al.}(2014)Regunathan, Sun, Tu, and
  Lin]{regunathan2014adaptive}
Regunathan, S.; Sun, S.; Tu, C.; Lin, C.L.
\newblock Adaptive quantization for enhancement layer video coding,  2014.
\newblock US Patent 8,897,359.

\bibitem[Soong \em{et~al.}(1987)Soong, Rosenberg, Juang, and
  Rabiner]{soong1987report}
Soong, F.K.; Rosenberg, A.E.; Juang, B.H.; Rabiner, L.R.
\newblock Report: A vector quantization approach to speaker recognition.
\newblock {\em AT\&T technical journal} {\bf 1987}, {\em 66},~14--26.

\bibitem[Hasan \em{et~al.}(2004)Hasan, Jamil, Rahman, et~al.]{hasan2004speaker}
Hasan, M.R.; Jamil, M.; Rahman, M.; others.
\newblock Speaker identification using mel frequency cepstral coefficients.
\newblock {\em variations} {\bf 2004}, {\em 1}.

\bibitem[Wu and Chang(2005)]{wu2005novel}
Wu, H.C.; Chang, C.C.
\newblock A novel digital image watermarking scheme based on the vector
  quantization technique.
\newblock {\em Computers \& Security} {\bf 2005}, {\em 24},~460--471.

\bibitem[Lu \em{et~al.}(2005)Lu, Xu, and Sun]{lu2005multipurpose}
Lu, Z.M.; Xu, D.G.; Sun, S.H.
\newblock Multipurpose image watermarking algorithm based on multistage vector
  quantization.
\newblock {\em IEEE Transactions on Image Processing} {\bf 2005}, {\em
  14},~822--831.

\bibitem[Equitz(1989)]{equitz1989new}
Equitz, W.H.
\newblock A new vector quantization clustering algorithm.
\newblock {\em IEEE transactions on acoustics, speech, and signal processing}
  {\bf 1989}, {\em 37},~1568--1575.

\bibitem[Lughofer(2008)]{lughofer2008extensions}
Lughofer, E.
\newblock Extensions of vector quantization for incremental clustering.
\newblock {\em Pattern recognition} {\bf 2008}, {\em 41},~995--1011.

\bibitem[Berger(2003)]{berger2003rate}
Berger, T.
\newblock Rate-distortion theory.
\newblock {\em Wiley Encyclopedia of Telecommunications} {\bf 2003}.

\bibitem[Shannon(1948)]{shannon1948mathematical}
Shannon, C.E.
\newblock A mathematical theory of communication.
\newblock {\em The Bell system technical journal} {\bf 1948}, {\em
  27},~379--423.

\bibitem[Mezghani and Nossek(2007)]{mezghani2007ultra}
Mezghani, A.; Nossek, J.A.
\newblock On ultra-wideband MIMO systems with 1-bit quantized outputs:
  Performance analysis and input optimization.
\newblock  2007 IEEE International Symposium on Information Theory. IEEE,
  2007, pp. 1286--1289.

\bibitem[Sundstrom \em{et~al.}(2008)Sundstrom, Murmann, and
  Svensson]{sundstrom2008power}
Sundstrom, T.; Murmann, B.; Svensson, C.
\newblock Power dissipation bounds for high-speed Nyquist analog-to-digital
  converters.
\newblock {\em IEEE Transactions on Circuits and Systems I: Regular Papers}
  {\bf 2008}, {\em 56},~509--518.

\bibitem[Mo and Heath(2014)]{mo2014high}
Mo, J.; Heath, R.W.
\newblock High {SNR} capacity of millimeter wave {MIMO} systems with one-bit
  quantization.
\newblock  2014 Information Theory and Applications Workshop (ITA). IEEE,
  2014, pp. 1--5.

\bibitem[Rini \em{et~al.}(2017)Rini, Barletta, Eldar, and
  Erkip]{rini2017general}
Rini, S.; Barletta, L.; Eldar, Y.C.; Erkip, E.
\newblock A general framework for {MIMO} receivers with low-resolution
  quantization.
\newblock  2017 IEEE Information Theory Workshop (ITW). IEEE,  2017, pp.
  599--603.

\bibitem[Khalili \em{et~al.}(2018)Khalili, Rini, Barletta, Erkip, and
  Eldar]{khalili2018mimo}
Khalili, A.; Rini, S.; Barletta, L.; Erkip, E.; Eldar, Y.C.
\newblock On {MIMO} channel capacity with output quantization constraints.
\newblock  2018 IEEE International Symposium on Information Theory (ISIT).
  IEEE,  2018, pp. 1355--1359.

\bibitem[Shlezinger \em{et~al.}(2018)Shlezinger, Eldar, and
  Rodrigues]{shlezinger2018hardware}
Shlezinger, N.; Eldar, Y.C.; Rodrigues, M.R.
\newblock Hardware-limited task-based quantization.
\newblock {\em arXiv preprint arXiv:1807.08305} {\bf 2018}.

\bibitem[Shlezinger and Eldar(2019)]{shlezinger2019deeptask}
Shlezinger, N.; Eldar, Y.C.
\newblock Deep task-based quantization.
\newblock {\em arXiv preprint arXiv:1908.06845} {\bf 2019}.

\bibitem[{Shohat} \em{et~al.}(2019){Shohat}, {Tsintsadze}, {Shlezinger}, and
  {Eldar}]{eldar2019deepquantization}
{Shohat}, M.; {Tsintsadze}, G.; {Shlezinger}, N.; {Eldar}, Y.C.
\newblock Deep Quantization for MIMO Channel Estimation.
\newblock  ICASSP 2019 - 2019 IEEE International Conference on Acoustics,
  Speech and Signal Processing (ICASSP),  2019, pp. 3912--3916.
\newblock
  doi:{\changeurlcolor{black}\href{https://doi.org/10.1109/ICASSP.2019.8682704}{\detokenize{10.1109/ICASSP.2019.8682704}}}.

\bibitem[{Khobahi} \em{et~al.}(2019){Khobahi}, {Naimipour}, {Soltanalian}, and
  {Eldar}]{eldar2019deeprecovery}
{Khobahi}, S.; {Naimipour}, N.; {Soltanalian}, M.; {Eldar}, Y.C.
\newblock Deep Signal Recovery with One-bit Quantization.
\newblock  ICASSP 2019 - 2019 IEEE International Conference on Acoustics,
  Speech and Signal Processing (ICASSP),  2019, pp. 2987--2991.
\newblock
  doi:{\changeurlcolor{black}\href{https://doi.org/10.1109/ICASSP.2019.8683876}{\detokenize{10.1109/ICASSP.2019.8683876}}}.

\bibitem[Shlezinger \em{et~al.}(2019)Shlezinger, Eldar, Salamatian, and
  Médard]{shlezinger2019quadratic}
Shlezinger, N.; Eldar, Y.C.; Salamatian, S.; Médard, M.
\newblock Task-based quantization for recovering quadratic functions using
  principal inertia components {\bf 2019}.

\bibitem[{Shlezinger} \em{et~al.}(2019){Shlezinger}, {Eldar}, and
  {Rodrigues}]{shlezinger2019asymptotic}
{Shlezinger}, N.; {Eldar}, Y.C.; {Rodrigues}, M.R.D.
\newblock Asymptotic Task-Based Quantization With Application to Massive MIMO.
\newblock {\em IEEE Transactions on Signal Processing} {\bf 2019}, {\em
  67},~3995--4012.
\newblock
  doi:{\changeurlcolor{black}\href{https://doi.org/10.1109/TSP.2019.2923149}{\detokenize{10.1109/TSP.2019.2923149}}}.

\bibitem[Kamilov \em{et~al.}(2012)Kamilov, Bourquard, Amini, and
  Unser]{kamilov2012one}
Kamilov, U.S.; Bourquard, A.; Amini, A.; Unser, M.
\newblock One-bit measurements with adaptive thresholds.
\newblock {\em IEEE Signal Processing Letters} {\bf 2012}, {\em 19},~607--610.

\bibitem[Plan and Vershynin(2013)]{plan2013one}
Plan, Y.; Vershynin, R.
\newblock One-Bit Compressed Sensing by Linear Programming.
\newblock {\em Communications on Pure and Applied Mathematics} {\bf 2013}, {\em
  66},~1275--1297.

\bibitem[Khobahi \em{et~al.}(2019)Khobahi, Naimipour, Soltanalian, and
  Eldar]{khobahi2019deep}
Khobahi, S.; Naimipour, N.; Soltanalian, M.; Eldar, Y.C.
\newblock Deep signal recovery with one-bit quantization.
\newblock  ICASSP 2019-2019 IEEE International Conference on Acoustics, Speech
  and Signal Processing (ICASSP). IEEE,  2019, pp. 2987--2991.

\bibitem[Max(1960)]{max1960quantizing}
Max, S.
\newblock Quantizing for minimum distortion.
\newblock {\em IEEE transactions on information theory} {\bf 1960}, {\em
  6},~7--12.

\bibitem[Lloyd(1982)]{lloyd1982least}
Lloyd, S.
\newblock Least squares quantization in {PCM}.
\newblock {\em IEEE transactions on information theory} {\bf 1982}, {\em
  28},~129--137.

\bibitem[Linde(1980)]{lindebuzogray1980vector}
Linde, Buzo, G.
\newblock An Algorithm for Vector Quantizer Design.
\newblock {\em IEEE transactions on Communications} {\bf 1980}, {\em
  28},~84--95.

\bibitem[Cover(1999)]{cover1999elements}
Cover, T.M.
\newblock {\em Elements of information theory}; John Wiley \& Sons,  1999.

\bibitem[Conway and Sloane(1982)]{conway1982fast}
Conway, J.; Sloane, N.
\newblock Fast quantizing and decoding and algorithms for lattice quantizers
  and codes.
\newblock {\em IEEE Transactions on Information Theory} {\bf 1982}, {\em
  28},~227--232.

\bibitem[Zamir and Feder(1992)]{zamir1992universal}
Zamir, R.; Feder, M.
\newblock On universal quantization by randomized uniform/lattice quantizers.
\newblock {\em IEEE Transactions on Information Theory} {\bf 1992}, {\em
  38},~428--436.

\bibitem[Stanley(2006)]{stanley2006}
Stanley, R.P.
\newblock {\em An introduction to Hyperplane Arrangements};  2006.

\bibitem[Dimca and PARUSINSKI(2017)]{dimca2017hyperplane}
Dimca, A.; PARUSINSKI, A.
\newblock {\em Hyperplane arrangements}; Springer,  2017.

\bibitem[Nelson and Gailly(1995)]{nelson1995data}
Nelson, M.; Gailly, J.L.
\newblock The data compression book 2nd edition.
\newblock {\em M \& T Books, New York, NY} {\bf 1995}.

\bibitem[Chou \em{et~al.}(1989)Chou, Lookabaugh, and Gray]{chou1989entropy}
Chou, P.A.; Lookabaugh, T.; Gray, R.M.
\newblock Entropy-constrained vector quantization.
\newblock {\em IEEE Transactions on Acoustics, Speech, and Signal Processing}
  {\bf 1989}, {\em 37},~31--42.

\bibitem[Orlik and Terao(2013)]{orlik2013arrangements}
Orlik, P.; Terao, H.
\newblock {\em Arrangements of hyperplanes}; Vol. 300, Springer Science \&
  Business Media,  2013.

\bibitem[Chataignon()]{coderepo}
Chataignon, J.
\newblock Comparison-limited vector quantizer design algorithm.
\newblock
  doi:{\changeurlcolor{black}\href{https://doi.org/10.5281/zenodo.3552797}{\detokenize{10.5281/zenodo.3552797}}}.

\bibitem[Bjorner \em{et~al.}(1994)Bjorner, Las~Vergnas, Sturmfels, White, and
  Ziegler]{bjorner1994}
Bjorner, A.; Las~Vergnas, M.; Sturmfels, B.; White, N.; Ziegler, G.
\newblock {\em Oriented Matroids};  1994.

\bibitem[Vakil(2006)]{vakil2006murphy}
Vakil, R.
\newblock Murphy’s law in algebraic geometry: Badly-behaved deformation
  spaces.
\newblock {\em Inventiones mathematicae} {\bf 2006}, {\em 164},~569--590.

\bibitem[Mnëv(1986)]{mnev1986}
Mnëv, N.
\newblock The topology of configuration varieties and convex polytopes
  varieties.
\newblock PhD thesis, Stanford University,  1986.

\end{thebibliography}






\end{document}